\documentclass[12pt,prd,showpacs,tightenlines,nofootinbib]{revtex4}
\usepackage{bm}
\usepackage{graphics}
\usepackage{rotating}
\usepackage{epsfig}
\begin{document}
\title{Charmless weak $B_s$ decays in the
  relativistic quark model } 

\author{R. N. Faustov}
\author{V. O. Galkin}
\affiliation{Dorodnicyn Computing Centre, Russian Academy of Sciences,
  Vavilov Str. 40, 119333 Moscow, Russia}

\begin{abstract}
The form factors of the weak $B_s$ transitions to the ground state and
orbitally excited strange mesons are calculated in the framework
of the QCD-motivated relativistic quark model based on the
quasipotential approach. These form factors are expressed through the
overlap integrals of meson wave functions found in their mass
spectrum evaluations. The momentum dependence of the form factors
is determined in the whole accessible kinematical range without any
additional assumptions and extrapolations. Relativistic effects,
including the wave function transformation from the rest to a moving reference
frame as well as contributions of the intermediate negative-energy
states, are consistently taken into account. The calculated form factors
are used for the evaluation of the charmless semileptonic decay rates  and two-body
nonleptonic $B_s$ decays in the factorization approximation. The
obtained results are confronted with previous predictions and
available experimental data. 
\end{abstract}


\maketitle

\section{Introduction}
\label{sec:int}

In last years the study of the properties of $B_s$ mesons and
especially its weak decays attracts attention of both theorists and
experimenters. Indeed the investigation of such decays is important for the
independent determination of the Cabibbo-Kobayashi-Maskawa (CKM)
matrix elements, studying the $CP$-violation, testing the standard
model and a ``new physics'' models. New precise data are coming
from the experiments on Large Hadron Collider (LHC) which significantly extended the
number of the observed $B_s$ decay channels \cite{lhcb,pdg}.

In the recent paper \cite{bsdecay} we considered the weak $B_s$ decays to
the ground and excited states of charmed mesons. Such decays are the
dominant decay channels of the $B_s$ meson. We calculated the weak
decay form factors in the framework of the relativistic quark model
with the QCD-motivated interquark potential. Good agreement of our
predictions for the semileptonic and nonleptonic decay branching
fractions with experimental data has been found. Here we extend our analysis to the
consideration of the CKM suppressed $B_s$ decays to strange
mesons. First we calculate the corresponding weak decay form
factors. The characteristic features of the heavy-to-light $B_s\to K$
transitions are the presence of the light $K$ meson in the final state and
a very broad accessible kinematical region. Therefore it is important
to consistently take into account  the relativistic effects, which give
substantial contributions, and to reliably determine the momentum
dependence of the form factors. It is
necessary to point out that most of the available theoretical
approaches permit to evaluate the form factors
either at some fixed point (of zero or maximum recoil of a final light
meson) or  determine the momentum dependence of  the form factors in a restricted kinematical range. As a result they require
ad hoc assumptions about the $q^2$ dependence of  form factors or
their extrapolations, which also rely on specific
models. The advantage of our model consists in its ability to
explicitly determine the corresponding decay form factors as the overlap
integrals of meson wave functions in the whole kinematical range without any
additional assumptions and extrapolations.    

In our calculations of the decay form factors we use the wave
functions of the $B_s$ and $K$ mesons obtained previously in their mass
spectra evaluations \cite{hlm,lm}. In Table~\ref{tab:smm} we
compare our predictions for the masses of the ground state and lowest
orbitally excited strange mesons \cite{lm} with available experimental
data \cite{pdg}. We find satisfactory agreement of our predictions with
data. Note that in our mass spectrum calculations all relativistic
effects, including the spin-dependent and spin-independent contributions to the potential, were treated
nonperturbatively in $v^2/c^2$. From Table~\ref{tab:smm} we see that
our model predicts that all orbitally excited $K^{(*)}_J$ mesons have
masses heavier than 1~GeV. The scalar $K^{*}_0(800)$ (or $\kappa$)
meson is predicted in our model to be a scalar tetraquark
\cite{ltetr}, while some other theoretical approaches assume it to be
the scalar quark-antiquark ($1^3P_0$) state. It is clear
that the form factors of the weak $B_s\to K^*_0$ transition are significantly
different if the $K^*_0(1430)$ meson is the $1^3P_0$ or $2^3P_0$ state. Therefore the resulting
rates of the semileptonic and nonleptonic $B_s$ decays to the $K^*_0(1430)$
strongly vary depending on its structure and quantum numbers. Thus the study of the weak
$B_s$ decays to the $K^*_0$ meson can help to reveal the nature of
light scalar mesons.  In the following we denote the
$K^*_0(1430)$ meson by $K^*_0$.     

\begin{table}
  \caption{Masses of the ground state and first orbitally excited strange
    mesons calculates in our model (in MeV).} 
  \label{tab:smm}
\begin{ruledtabular}
\begin{tabular}{ccccc}
$n^{2S+1}L_J$&$J^{P}$& Meson& Theory \cite{lm}&Experiment \cite{pdg}\\
\hline
$1^1S_0$& $0^{-}$& $K$&482&493.677(16)\\
$1^3S_1$& $1^{-}$& $K^*(892)$&897&891.66(26)\\
$1^3P_0$& $0^{+}$& $K_0^*(1430)$&1362&1425(50)\\
$1P_1$& $1^{+}$& $K_1(1270)$&1294&1272(7)\\
$1P_1$& $1^{+}$& $K_1(1400)$&1412&1403(7)\\
$1^3P_2$& $2^{+}$& $K_2^*(1430)$&1424&1425.6(15)\\
\end{tabular}
 \end{ruledtabular}
\end{table}

The calculated form factors are used for evaluation of the rates of the
charmless semileptonic decays both to the ground state and
orbitally excited strange mesons. The two-body tree-dominated
nonleptonic $B_s$ decays are considered within the factorization
approximation. The charmless $B_s$ decay rates are evaluated and
compared with previous calculations and available experimental data.

\section{Relativistic quark model}  
\label{rqm}

The employed relativistic quark model is based on the
quasipotential approach in quantum chromodynamics (QCD).  Hadrons are
considered as the bound states of constituent quarks which are described by the
single-time wave functions satisfying the
three-dimensional relativistically invariant Schr\"odinger-like
equation with the QCD-motivated interquark potential \cite{mass}  
\begin{equation}
\label{quas}
{\left(\frac{b^2(M)}{2\mu_{R}}-\frac{{\bf
p}^2}{2\mu_{R}}\right)\Psi_{M}({\bf p})} =\int\frac{d^3 q}{(2\pi)^3}
 V({\bf p,q};M)\Psi_{M}({\bf q}),
\end{equation}
where the relativistic reduced mass is
\begin{equation}
\mu_{R}=\frac{M^4-(m^2_1-m^2_2)^2}{4M^3},
\end{equation}
 $M$ is the meson mass, $m_{1,2}$ are the quark masses,
and ${\bf p}$ is their relative momentum.  
In the center of mass system the relative momentum squared on mass shell 
$b^2(M)$ is expressed through the meson and quark masses:
\begin{equation}
{b^2(M) }
=\frac{[M^2-(m_1+m_2)^2][M^2-(m_1-m_2)^2]}{4M^2}.
\end{equation}
The kernel of this equation is the interquark quasipotential $V({\bf
  p,q};M)$ which consists of the perturbative one-gluon
exchange and the nonperturbative confining parts \cite{mass}
  \begin{equation}
\label{qpot}
V({\bf p,q};M)=\bar{u}_1(p)\bar{u}_2(-p){\mathcal V}({\bf p}, {\bf
q};M)u_1(q)u_2(-q), 
\end{equation}
with
$${\mathcal V}({\bf p},{\bf q};M)=\frac{4}{3}\alpha_sD_{ \mu\nu}({\bf
k})\gamma_1^{\mu}\gamma_2^{\nu}
+V^V_{\rm conf}({\bf k})\Gamma_1^{\mu}({\bf k})
\Gamma_{2;\mu}({\bf k})+V^S_{\rm conf}({\bf k}),\qquad {\bf k=p-q},$$
where $\alpha_s$ is the QCD coupling constant, $D_{\mu\nu}$ is the
gluon propagator in the Coulomb gauge, and $\gamma_{\mu}$ and $u(p)$ are  the Dirac matrices and
spinors, respectively. The Lorentz
structure of the confining part includes the scalar and vector linearly
rising interactions which in the nonrelativistic limit reduce to
\begin{equation}
\label{nr}
V_{\rm conf}(r)=V_{\rm conf}^S(r)+V_{\rm conf}^V(r)=Ar+B,
\end{equation}
with
\begin{equation}
\label{vlin}
V_{\rm conf}^V(r)=(1-\varepsilon)(Ar+B),\qquad 
V_{\rm conf}^S(r) =\varepsilon (Ar+B),
\end{equation}
where $\varepsilon$ is the mixing coefficient. Its value  $\varepsilon=-1$
has been obtained from the consideration of the heavy quark expansion
for the semileptonic $B\to D$ decays
\cite{fg} and charmonium radiative decays \cite{mass}.

The long-range vector vertex 
\begin{equation}
\label{kappa}
\Gamma_{\mu}({\bf k})=\gamma_{\mu}+
\frac{i\kappa}{2m}\sigma_{\mu\nu}k^{\nu}
\end{equation}
contains the Pauli
term with anomalous chromomagnetic quark moment $\kappa$. The  value  $\kappa=-1$,
fixed in our model from the analysis of the fine splitting of heavy quarkonia ${
}^3P_J$- states \cite{mass} and  the heavy quark expansion for semileptonic
decays of heavy mesons \cite{fg} and baryons \cite{sbar},
enables vanishing of the spin-dependent chromomagnetic interaction,
proportional to $(1+\kappa)$, in
accord with the flux tube model. 

Other parameters of our model  were determined from the previous analysis of
meson spectroscopy  \cite{mass}. The
constituent quark masses
are $m_b=4.88$ GeV, $m_c=1.55$ GeV, $m_s=0.5$ GeV, $m_{u,d}=0.33$ GeV and
the parameters of the linear potential are $A=0.18$ GeV$^2$ and
$B=-0.30$ GeV.  

For the consideration of the meson weak decays it is necessary to
calculate the matrix element of the weak current between meson
states. In the quasipotential approach such matrix element between a $B_s$ meson with mass $M_{B_s}$ and
momentum $p_{B_s}$ and a final $K$ meson with mass $M_{K}$ and
momentum $p_{K}$ is given by  \cite{f}
\begin{equation}\label{mxet} 
\langle K(p_{K}) \vert J^W_\mu \vert B_s(p_{B_s})\rangle
=\int \frac{d^3p\, d^3q}{(2\pi )^6} \bar \Psi_{{K}\,{\bf p}_K}({\bf
p})\Gamma _\mu ({\bf p},{\bf q})\Psi_{B_s\,{\bf p}_{B_s}}({\bf q}),
\end{equation}
where $\Gamma _\mu ({\bf p},{\bf
q})$ is the two-particle vertex function and  
$\Psi_{M\,{\bf p}_M}({\bf p})$ are the
meson ($M=B_s,{K})$ wave functions projected onto the positive energy states of
quarks and boosted to the moving reference frame with momentum ${\bf
  p}_M$, and  ${\bf p},{\bf q}$ are relative quark momenta. 

It is convenient to carry calculations in the $B_s$ meson
rest frame (${\bf p}_{B_s}=0$). Then the final  meson
is moving with the recoil momentum ${\bf \Delta}$. The wave function
of the moving  meson $\Psi_{K\,{\bf\Delta}}$ is connected 
with the  wave function in the rest frame 
$\Psi_{K\,{\bf 0}}\equiv \Psi_{K}$ by the transformation \cite{f}
\begin{equation}
\label{wig}
\Psi_{K\,{\bf\Delta}}({\bf
p})=D_u^{1/2}(R_{L_{\bf\Delta}}^W)D_s^{1/2}(R_{L_{
\bf\Delta}}^W)\Psi_{K\,{\bf 0}}({\bf p}),
\end{equation}
where $R^W$ is the Wigner rotation, $L_{\bf\Delta}$ is the Lorentz boost
from the meson rest frame to a moving one and $D^{1/2}(R)$ is  
the spin rotation matrix.

The wave function of a final $^{2S+1}K_{J}$ meson at rest is given by
\begin{equation}\label{psi}
\Psi_{K}({\bf p})\equiv
\Psi^{JLS{\cal M}}_{^{2S+1}K_{J}}({\bf p})={\cal Y}^{JLS{\cal M}}\,\psi_{^{2S+1}K_{J}}({\bf p}),
\end{equation}
where $J$ and ${\cal M}$ are the total meson angular momentum and its
projection,  $L$ is the orbital momentum,
while $S=0,1$ is the total spin.   
$\psi_{^{2S+1}K_{J}}({\bf p})$ is the radial part of the wave function.
The spin-angular momentum part ${\cal Y}^{JLS{\cal M}}$ is defined by
\begin{equation}\label{angl}
{\cal Y}^{JLS{\cal M}}=\sum_{\sigma_1\sigma_2}\langle L\, {\cal M}-\sigma_1-\sigma_2,\  
S\, \sigma_1+\sigma_2 |J\, {\cal M}\rangle\langle \frac12\, \sigma_1,\ 
\frac12\, \sigma_2 |S\, \sigma_1+\sigma_2\rangle Y_{L}^{{\cal M}-\sigma_1-\sigma_2}
\chi_1(\sigma_1)\chi_2(\sigma_2),
\end{equation}
where $\langle j_1\, m_1,\  j_2\, m_2|J\, {\cal M}\rangle$ are the Clebsch-Gordan 
coefficients, $Y_l^m$ are the spherical harmonics, and $\chi(\sigma)$ (where 
$\sigma=\pm 1/2$) are the spin wave functions.

\begin{figure}
  \centering
  \includegraphics{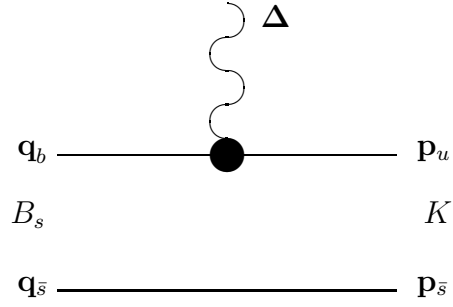}
\caption{Leading order vertex function $\Gamma^{(1)}({\bf p},{\bf q})$
contributing to the current matrix element (\ref{mxet}). \label{d1}}
\end{figure}

The explicit expression for the vertex function  $\Gamma_\mu ({\bf
  p},{\bf q})$ can be found in Ref.~\cite{bsdecay}. It contains
contributions both from the leading order spectator diagram (see Fig.~\ref{d1}) and from
subleading order diagrams accounting for the contributions of the
negative-energy intermediate states. The leading order contribution
\begin{equation} \label{gamma1}
\Gamma_\mu^{(1)}({\bf
p},{\bf q})=\bar u_{c}(p_c)\gamma_\mu(1-\gamma^5)u_b(q_b)
(2\pi)^3\delta({\bf p}_{\bar s}-{\bf
q}_{\bar s})\end{equation}
contains the $\delta$ function which allows us to take one of the
integrals in the matrix element (\ref{mxet}).  Calculation of the subleading order
contribution is more complicated due to the dependence on
the relative momentum in the energies of the initial heavy and final
light quarks. For the energy of the heavy quark we use heavy quark
expansion. For the light quark such expansion is not applicable.  However, 
the final light $K$ meson has a large (compared to its mass) recoil momentum
($|{\bf\Delta}_{\rm  max}|=(M_{B_s}^2-M_K^2)/(2M_{B_s})\sim 2.6$~GeV)  almost in
the whole kinematical range except the small region near  
$q^2=q^2_{\rm max}$ ($|{\bf\Delta}|=0$).  This also means that the
recoil momentum of the final meson is large with respect to the mean
relative quark momentum $|{\bf p}|$ in the meson  ($\sim 0.5$~GeV).
Thus one can neglect  $|{\bf p}|$ compared to $|{\bf\Delta}|$ in  the
light quark energies
$\epsilon_{q}(p+\Delta)\equiv\sqrt{m_{q}^2+({\bf 
p}+{\bf\Delta})^2}$, replacing it by  $\epsilon_{q}(\Delta)\equiv
\sqrt{m_{q}^2+{\bf\Delta}^2}$  in expressions for the
subleading contribution.  Such replacement removes the relative
momentum dependence in the energies of  quarks and thus permits
to perform one of the integrations in the subleading 
contribution using the quasipotential equation. Since the subleading
contributions are suppressed the uncertainty introduced by such
procedure is small.  As the result the weak
decay matrix element is expressed through the usual overlap integral
of initial and final meson wave functions and its momentum dependence
can be determined in the whole accessible kinematical range without
additional assumptions. 

\section{Form factors of the weak transitions of $B_s$ to $K$ mesons}
\label{sec:ffbsk}

The matrix elements of the  vector, axial vector and tensor weak
currents between $B_s$ and 
$K^{(*)}$ meson states are parametrized by the following set of form factors 
\begin{equation}
  \label{eq:pff1}
  \langle K(p_{K})|\bar u \gamma^\mu b|B_s(p_{B_s})\rangle
  =f_+(q^2)\left[p_{B_s}^\mu+ p_{K}^\mu-
\frac{M_{B_s}^2-M_{K}^2}{q^2}\ q^\mu\right]+
  f_0(q^2)\frac{M_{B_s}^2-M_{K}^2}{q^2}\ q^\mu,
\end{equation}
\begin{equation}
\label{eq:pff12} 
 \langle K(p_{K})|\bar u \gamma^\mu\gamma_5 b|B_s(p_{B_s})\rangle
  =0,
\end{equation}
\begin{equation}
\label{eq:pff2}
\langle K(p_K)|\bar u \sigma^{\mu\nu}q_\nu b|B_s(p_{B_s})\rangle=
\frac{if_T(q^2)}{M_{B_s}+M_K} [q^2(p_{B_s}^\mu+p_K^\mu)-(M_{B_s}^2-M_K^2)q^\mu],
\end{equation}
\begin{eqnarray}
  \label{eq:vff1}
  \langle {K^*}(p_{K^*})|\bar u \gamma^\mu b|B_s(p_{B_s})\rangle&=
  &\frac{2iV(q^2)}{M_{B_s}+M_{K^*}} \epsilon^{\mu\nu\rho\sigma}\epsilon^*_\nu
  p_{B_s\rho} p_{{K^*}\sigma},\\ \cr
\label{eq:vff2}
\langle {K^*}(p_{K^*})|\bar u \gamma^\mu\gamma_5 b|B_s(p_{B_s})\rangle&=&2M_{K^*}
A_0(q^2)\frac{\epsilon^*\cdot q}{q^2}\ q^\mu
 +(M_{B_s}+M_{K^*})A_1(q^2)\left(\epsilon^{*\mu}-\frac{\epsilon^*\cdot
    q}{q^2}\ q^\mu\right)\cr\cr
&&-A_2(q^2)\frac{\epsilon^*\cdot q}{M_{B_s}+M_{K^*}}\left[p_{B_s}^\mu+
  p_{K^*}^\mu-\frac{M_{B_s}^2-M_{K^*}^2}{q^2}\ q^\mu\right], 
\end{eqnarray}
\begin{equation}
  \label{eq:vff3}
\langle  K^*(p_{K^*})|\bar u i\sigma^{\mu\nu}q_\nu b|B_s(p_{B_s})\rangle=2T_1(q^2)
\epsilon^{\mu\nu\rho\sigma} \epsilon^*_\nu p_{{K^*}\rho} p_{{B_s}\sigma},
\end{equation}
\begin{eqnarray}
\label{eq:vff4}
\langle K^*(p_{K^*})|\bar u i\sigma^{\mu\nu}\gamma_5q_\nu b|B_s(p_{B_s})\rangle&=&
T_2(q^2)[(M_{B_s}^2-M_{K^*}^2)\epsilon^{*\mu}-(\epsilon^*\cdot q)(p_{B_s}^\mu+
p_{K^*}^\mu)]\cr\cr
&&+T_3(q^2)(\epsilon^*\cdot q)\left[q^\mu-\frac{q^2}{M_{B_s}^2-M_{K^*}^2}
  (p_{B_s}^\mu+p_{K^*}^\mu)\right],
\end{eqnarray}
 $q=p_{B_s}- p_{K^{(*)}}$, and $M_{B,K^{(*)}}$ are the masses of the $B$ meson  
and final $K^{(*)}$ meson, respectively; while  $\epsilon_\mu$ is the
  polarization vector of the final vector  $K^{*}$ meson.

At the maximum recoil point ($q^2=0$) these form
factors satisfy the following conditions: 
\[f_+(0)=f_0(0),\]
\[A_0(0)=\frac{M_{B_s}+M_{K^*}}{2M_{K^*}}A_1(0)
-\frac{M_{B_s}-M_{K^*}}{2M_{K^*}}A_2(0),\]
\[T_1(0)=T_2(0).\]

Comparing these decompositions with the results of the calculations of
the weak current matrix element in our model, as described in the previous section, we determine the form
factors in the whole accessible kinematical range through the overlap
integrals of the meson wave functions. The explicit expressions are
given in Refs.~\cite{bcdecay,brare}. For the numerical evaluations of
the corresponding overlap integrals we use the quasipotential wave
functions of $B_s$ and $K^{(*)}$ mesons obtained in their mass spectra
calculations \cite{hlm,lm}.

We find that the weak $B_s\to K^{(*)}$ transition form factors can 
be approximated with good accuracy by the following expressions
\cite{ms,bdecays}: 

(a) $F(q^2)= \{f_+(q^2),f_T(q^2),V(q^2),A_0(q^2),T_1(q^2)\}$ 
\begin{equation}
  \label{fitfv}
  F(q^2)=\frac{F(0)}{\displaystyle\left(1-\frac{q^2}{M^2}\right)
    \left(1-\sigma_1 
      \frac{q^2}{M_{B^*}^2}+ \sigma_2\frac{q^4}{M_{B^*}^4}\right)},
\end{equation}

\begin{figure}
\centering
  \includegraphics[width=8cm]{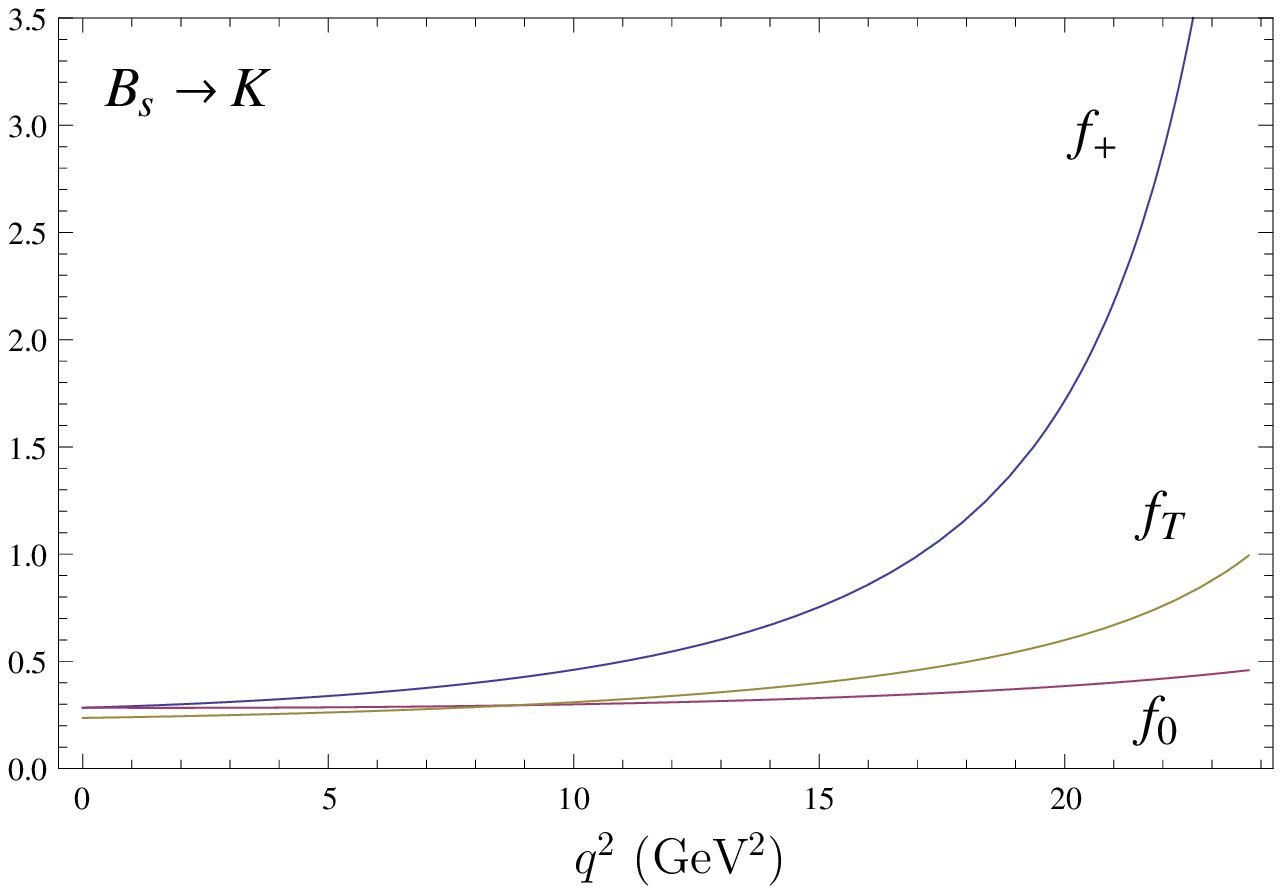}\ \
\caption{Form factors of the weak $B_s\to K$ transition.    } 
\label{fig:ffk}
\end{figure}

\begin{figure}
\centering
  \includegraphics[width=8cm]{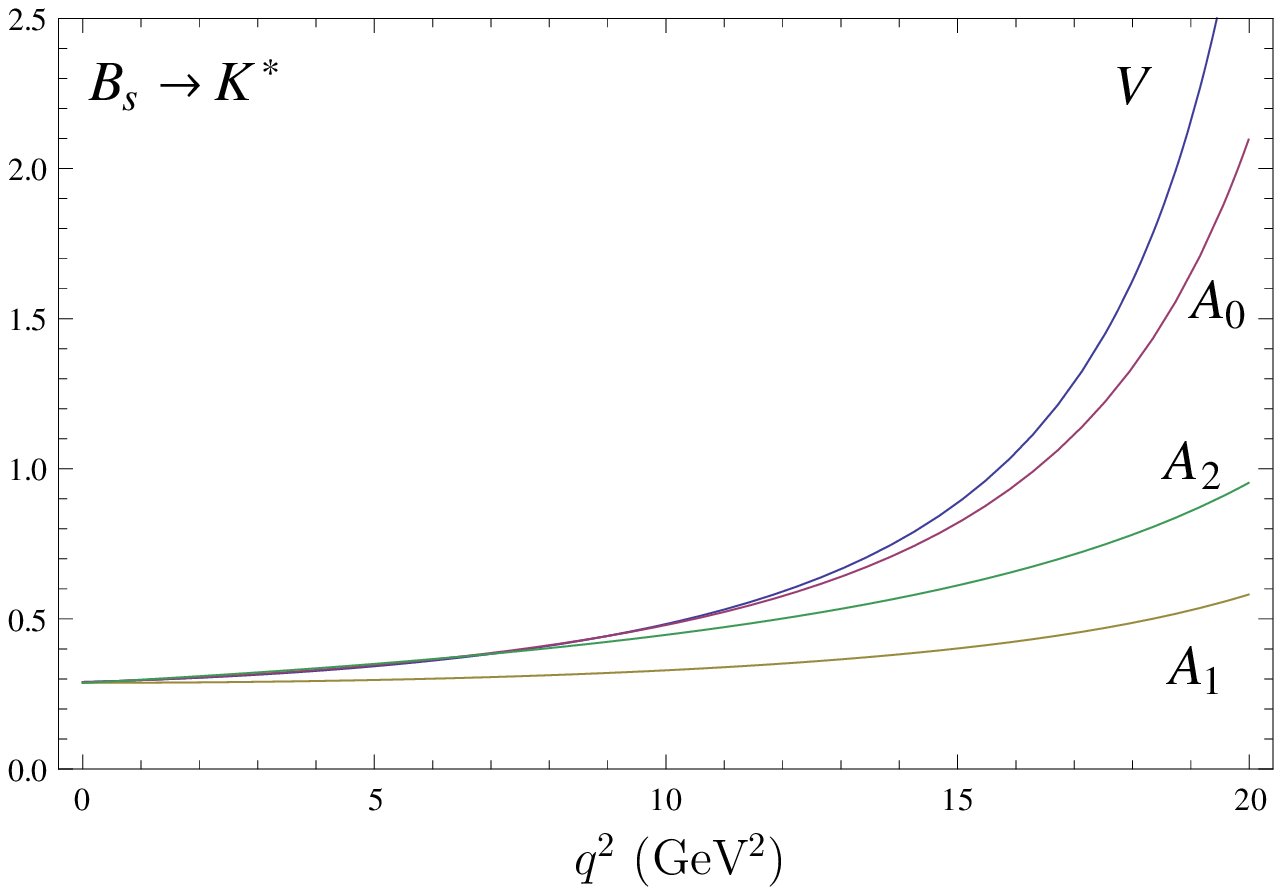}\ \
 \ \includegraphics[width=8cm]{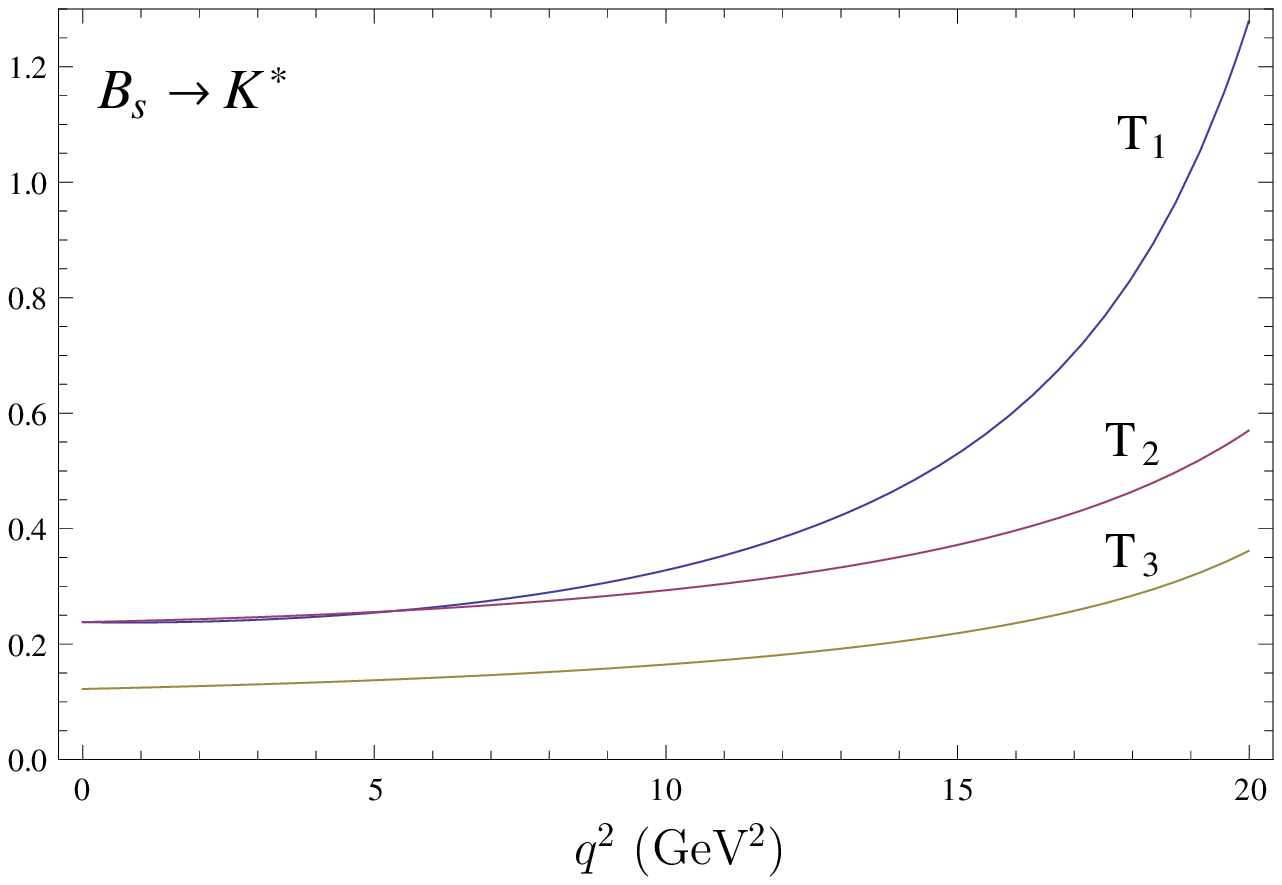}\\
\caption{Form factors of the weak $B_s\to K^{*}$ transition.    } 
\label{fig:ffks}
\end{figure}

(b) $F(q^2)=\{f_0(q^2), A_1(q^2),A_2(q^2),T_2(q^2),T_3(q^2)\}$
\begin{equation}
  \label{fita12}
  F(q^2)=\frac{F(0)}{\displaystyle \left(1-\sigma_1
      \frac{q^2}{M_{B^*}^2}+ \sigma_2\frac{q^4}{M_{B^*}^4}\right)},
\end{equation}
where $M=M_{B^*}$ for the form factors $f_+(q^2),f_T(q^2),V(q^2),T_1(q^2)$ and
$M=M_{B}$ for the form factor $A_0(q^2)$. The obtained values $F(0)$ and
$\sigma_{1,2}$ are given in Table~\ref{hff}. The difference between
fitted and calculated form factors is less than 1\%. We can roughly
estimate the total uncertainty of the form factors within
our model to be less than 5\%. It mainly originates from the subleading contributions to the decay
matrix elements in the region of small recoils. These form factors are
plotted in Figs.~\ref{fig:ffk}  and \ref{fig:ffks}.

\begin{table}
\caption{Calculated form factors of weak $B_s\to K^{(*)}$ transitions. Form factors $f_+(q^2)$, $f_T(q^2)$ $V(q^2)$,
  $A_0(q^2)$, $T_1(q^2)$ are fitted by Eq.~(\ref{fitfv}), and form factors $f_0(q^2)$,
  $A_1(q^2)$, $A_2(q^2)$, $T_2(q^2)$, $T_3(q^2)$  are fitted by Eq.~(\ref{fita12}).  }
\label{hff}
\begin{ruledtabular}
\begin{tabular}{ccccccccccc}
   &\multicolumn{3}{c}{{$B_s\to K$}}&\multicolumn{7}{c}{{\  $B_s\to K^*$\
     }}\\
\cline{2-4} \cline{5-11}
& $f_+$ & $f_0$& $f_T$& $V$ & $A_0$ &$A_1$&$A_2$& $T_1$ & $T_2$& $T_3$\\
\hline
$F(0)$          &0.284 &0.284 &  0.236 & 0.291 & 0.289& 0.287 & 0.286 & 0.238& 0.238& 0.122\\
$F(q^2_{\rm max})$&5.42  &0.459 &  0.993 & 3.06& 2.10& 0.581 &  0.953 & 1.28& 0.570& 0.362\\
$\sigma_1$      &$-0.370$&$-0.072$& $-0.442$& $-0.516$ &$-0.383$&  0&
1.05  &$-1.20$&$0.241$& $0.521$\\
$\sigma_2$      &$-1.41$&$-0.651$&$0.082$&$-2.10$&$-1.58$&$-1.06$&
0.074  &$-2.44$&$-0.857$& $-0.613$\\
\end{tabular}
\end{ruledtabular}
\end{table}

In Table~\ref{compbpiff} we compare our predictions for the form
factors of weak $B_s$ decays at maximum recoil  $q^2=0$ with
results of other calculations
\cite{bz,akllsww,ms,llw,lww,wzz,swyz,wx}. Light-cone sum rules, including
one-loop radiative corrections to twist-2 and twist-3 contributions,
and leading order twist-4 corrections are used in
Ref.~\cite{bz}. Perturbative QCD approach is applied in
Refs.~\cite{akllsww,llw}. Calculations based on quark model and
relativistic dispersion approach are given in Ref.~\cite{ms}, while
consideration in Ref.~\cite{lww} is performed in the light cone quark
model utilizing the soft collinear effective theory. The authors of
Ref.~\cite{wzz} employ light-cone sum rules in the framework of heavy
quark effective theory. In Ref.~\cite{swyz} the weak transition form
factors are evaluated in the six-quark effective Hamiltonian
approach. Perturbative QCD factorization approach with the inclusion
of the next-to-leading-order corrections is used in Ref.~\cite{wx}.
We find a reasonable agreement between the values of the weak $B_s\to
K^{(*)}$ transition form factors at $q^2=0$ calculated in
significantly different approaches.

\begin{sidewaystable}
\caption{Comparison of theoretical predictions for the form factors of 
  weak $B_s\to  K^{(*)}$ transitions at maximum
  recoil point $q^2=0$.  }
\label{compbpiff}
\begin{ruledtabular}
\begin{tabular}{ccccccccc}
     & $f_+(0)$& $f_T(0)$ & $V(0)$ & $A_0(0)$ &$A_1(0)$&$A_2(0)$ &$T_1(0)$&$T_3(0)$ \\
\hline
this paper   &$0.284\pm0.014$  & $0.236\pm0.012$  &$0.291\pm0.015$  &$0.289\pm0.015$ &$0.287\pm0.015$  &$0.286\pm0.015$ & $0.238\pm0.012$& $0.122\pm0.006$\\
\cite{bz}   &$0.30\pm0.04$  &   &$0.311\pm0.026$ & $0.360\pm0.034$ & $0.233\pm0.022$& $0.181\pm0.025$& $0.260\pm0.024$& $0.136\pm0.016$\\
\cite{akllsww} &$0.24\pm0.05$& &$0.21\pm0.04$ &$0.25\pm0.05$
&$0.16\pm0.04$& & \\
\cite{ms} &$0.31$ &0.31 & 0.38 &0.37&0.29&0.26&0.32&0.23
  \\
\cite{llw} & & &$0.20\pm0.05$&$0.24^{+0.07}_{-0.05}$& $0.15^{+0.04}_{-0.03}$& $0.11\pm0.02$& $0.18\pm0.05$& $0.16\pm0.03$\\
\cite{lww} &0.290 &0.317 &0.323 &0.279&0.232 &0.210&0.271&0.165\\
\cite{wzz} &$0.296\pm0.018$ &$0.288\pm0.018$ &$0.285\pm0.013$
&$0.222\pm0.011$&$0.227\pm0.011$&$0.183\pm0.010$&$0.251\pm0.012$&$0.169\pm0.008$ \\
\cite{swyz} &$0.260^{+0.055}_{-0.032}$& 
&$0.227^{+0.064}_{-0.037}$&$0.280^{+0.090}_{-0.045}$&$0.178^{+0.047}_{-0.027}$
\\
\cite{wx} &$0.26^{+0.05}_{-0.04}$ 
&$0.28\pm0.05$&
\end{tabular}
\end{ruledtabular}
\end{sidewaystable}

\section{Form factors of weak transitions of $B_s$ mesons to orbitally excited $K_{J}^{(*)}$ mesons}
\label{sec:fforbexc}

Now we apply the same approach for the calculation of the form factors
of the weak $B_s$ decays to orbitally excited $K_{J}^{(*)}$ mesons.
The matrix elements of the weak current $J^W_\mu=\bar
u\gamma_\mu(1-\gamma_5)b$ for $B_s$ decays to orbitally
excited $P$-wave $K_J^{(*)}$ mesons can be parametrized by the
following set of invariant
form factors
\begin{eqnarray}
  \label{eq:sff1}
\langle K_{0}^*(p_{K_{0}})|\bar u \gamma^\mu b|B_s(p_{B_s})\rangle
  &=&0,\\ \cr
  \langle K_{0}^*(p_{K_{0}})|\bar u \gamma^\mu\gamma_5 b|B_s(p_{B_s})\rangle
  &=&r_+(q^2)\left(p_{B_s}^\mu+ p_{K_{0}}^\mu\right)+
  r_-(q^2)\left(p_{B_s}^\mu- p_{K_{0}}^\mu\right),\\ \cr
  \label{eq:avff1}
  \langle K_{1}(p_{K_{1}})|\bar u \gamma^\mu b|B_s(p_{B_s})\rangle\!\!&=&\!\!
  (M_{B_s}+M_{K_{1}})h_{V_1}(q^2)\epsilon^{*\mu}
  +[h_{V_2}(q^2)p_{B_s}^\mu+h_{V_3}(q^2)p_{K_{1}}^\mu]\frac{\epsilon^*\cdot q}{M_{B_s}} ,\qquad\\\cr
\label{eq:avff2}
\langle K_{1}(p_{K_{1}})|\bar u \gamma^\mu\gamma_5 b|B_s(p_{B_s})\rangle\!\!&=&\!\!
\frac{2ih_A(q^2)}{M_{B_s}+M_{K_{1}}} \epsilon^{\mu\nu\rho\sigma}\epsilon^*_\nu
  p_{B_s\rho} p_{{K_{1}}\sigma},  \\ \cr
  \label{eq:tff1}
  \langle K_{2}^*(p_{K_{2}})|\bar u \gamma^\mu b|B_s(p_{B_s})\rangle&=&
\frac{2it_V(q^2)}{M_{B_s}+M_{K_{2}}} \epsilon^{\mu\nu\rho\sigma}\epsilon^*_{\nu\alpha}
\frac{p_{B_s}^\alpha}{M_{B_s}}  p_{B_s\rho} p_{{K_{2}}\sigma},\\\cr
\label{eq:tff2}
\langle K_{2}^*(p_{K_{2}})|\bar u \gamma^\mu\gamma_5 b|B_s(p_{B_s})\rangle&=&
(M_{B_s}+M_{K_{2}})t_{A_1}(q^2)\epsilon^{*\mu\alpha}\frac{p_{B_s\alpha}}{M_{B_s}}\cr\cr
&&  +[t_{A_2}(q^2)p_{B_s}^\mu+t_{A_3}(q^2)p_{K_{2}}^\mu]\epsilon^*_{\alpha\beta}
\frac{p_{B_s}^\alpha p_{B_s}^\beta}{M_{B_s}^2} , 
\end{eqnarray}
where $q=p_{B_s}-p_{K_{J}}$, $M_{K_{J}}$ are $P$-wave $K$ meson
masses, $\epsilon^\mu$ and $\epsilon^{\mu\nu}$ are the polarization
vector and tensor of the vector $K_1\equiv K_1(1270)$ and tensor
$K^*_2$ mesons, respectively.  The matrix elements of the weak current for
$B_s$ decays to the axial vector $K_{1}(1400)$ meson are
obtained from Eqs.~(\ref{eq:avff1}), (\ref{eq:avff2}) by the replacement of the set of
form factors $h_i(q^2)$ by $g_i(q^2)$ ($i=V_1,V_2,V_3,A$).   

The $P$-wave $K$ meson states  with $J=L=1$
are the mixtures of spin-triplet ($^3P_1$)  and spin-singlet ($^1P_1$)
states:
\begin{eqnarray}
  \label{eq:mix}
  |K_1(1270)\rangle&=&|K(^1P_1)\rangle\cos\varphi+|K(^3P_1)\rangle\sin\varphi, \cr
 |K_1(1400)\rangle&=&-|K(^1P_1)\rangle\sin\varphi+|K(^3P_1)\rangle\cos\varphi,
\end{eqnarray}
where $\varphi$ is a mixing angle.
  Such mixing occurs due to the nondiagonal spin-orbit and
tensor terms in the spin-dependent part of the relativistic
quasipotential. The masses of  physical states are obtained 
by diagonalizing the mixing terms. The found value of the mixing angle, 
$\varphi=43.8^\circ$ \cite{lm}, implies that physical $K_1$ mesons are
nearly equal mixtures of the spin-singlet $K(^1P_1)$ and
spin-triplet $K(^3P_1)$ states in accord with the experimental
data \cite{pdg}.

We calculate the matrix elements of the weak current between the
initial $B_s$ meson and final orbitally excited $P$-wave $K_J^{(*)}$
meson using the procedure described above and compare the result with
the invariant decomposition (\ref{eq:sff1})-(\ref{eq:tff2}). In this
way we obtain expressions for the decay form factors in terms of
the overlap integrals of the initial and final meson wave functions which are
valid in the whole accessible kinematical range. Note that these form
factors account for the relativistic transformations of the meson wave
functions from the rest to a moving reference frame (\ref{wig}) and
relativistic contributions from the intermediate negative energy
states. The explicit expressions for these form factors can be
obtained from the corresponding formulas in Appendix of
Ref.~\cite{bcdecay} with obvious replacements.  

\begin{figure}
  \centering
  \includegraphics[width=7.9cm]{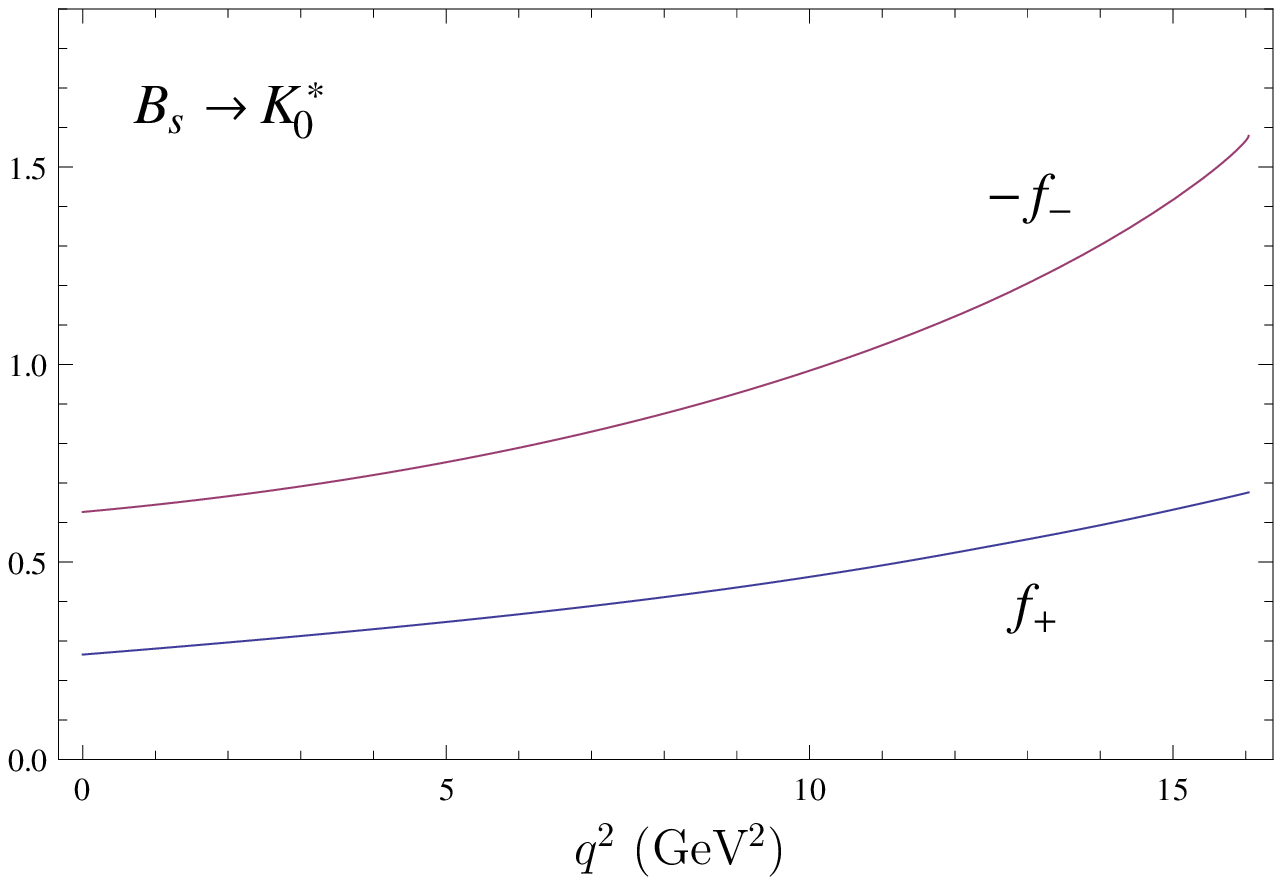} \ \ 
\  
\includegraphics[width=7.9cm]{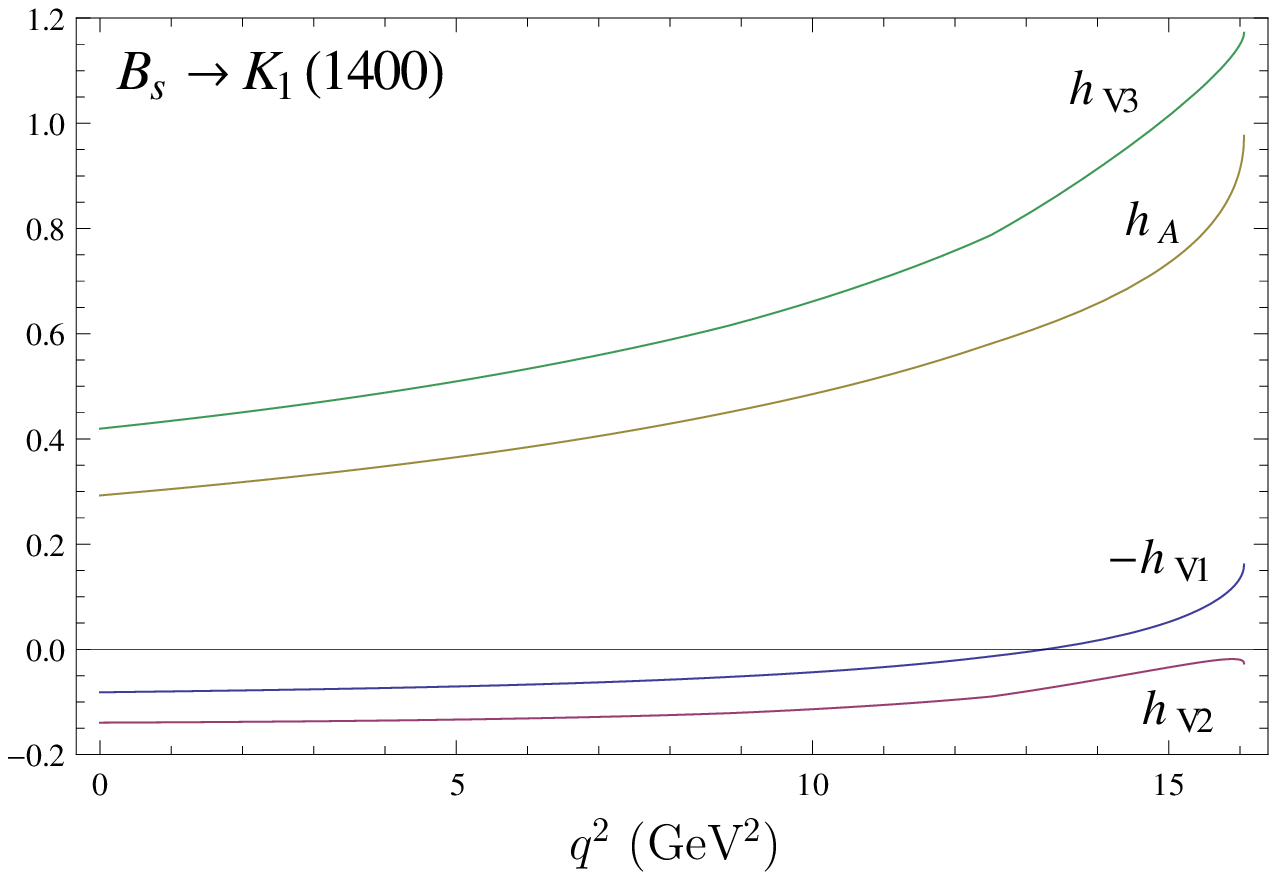}

\vspace*{0.5cm}

  \includegraphics[width=7.9cm]{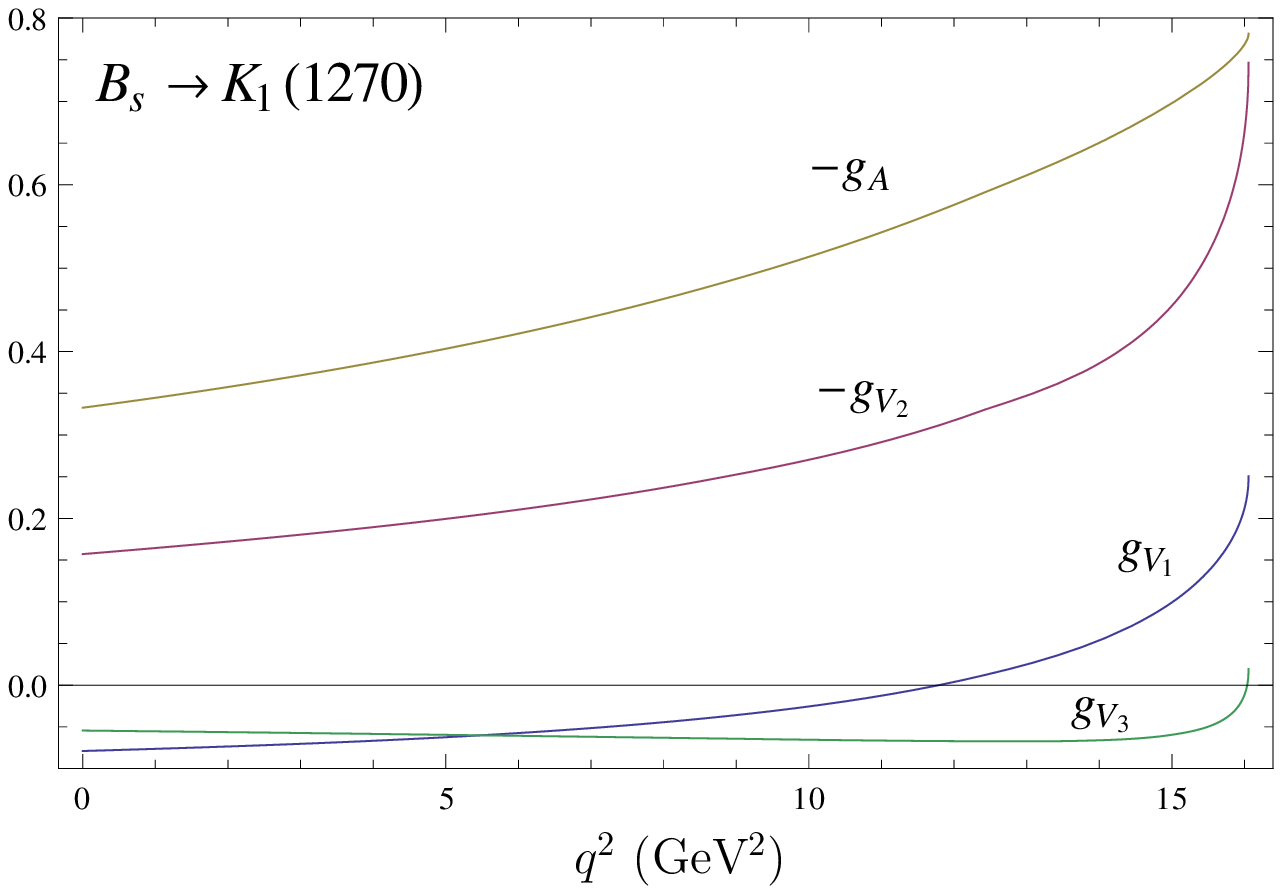} \ \ 
\  
\includegraphics[width=7.9cm]{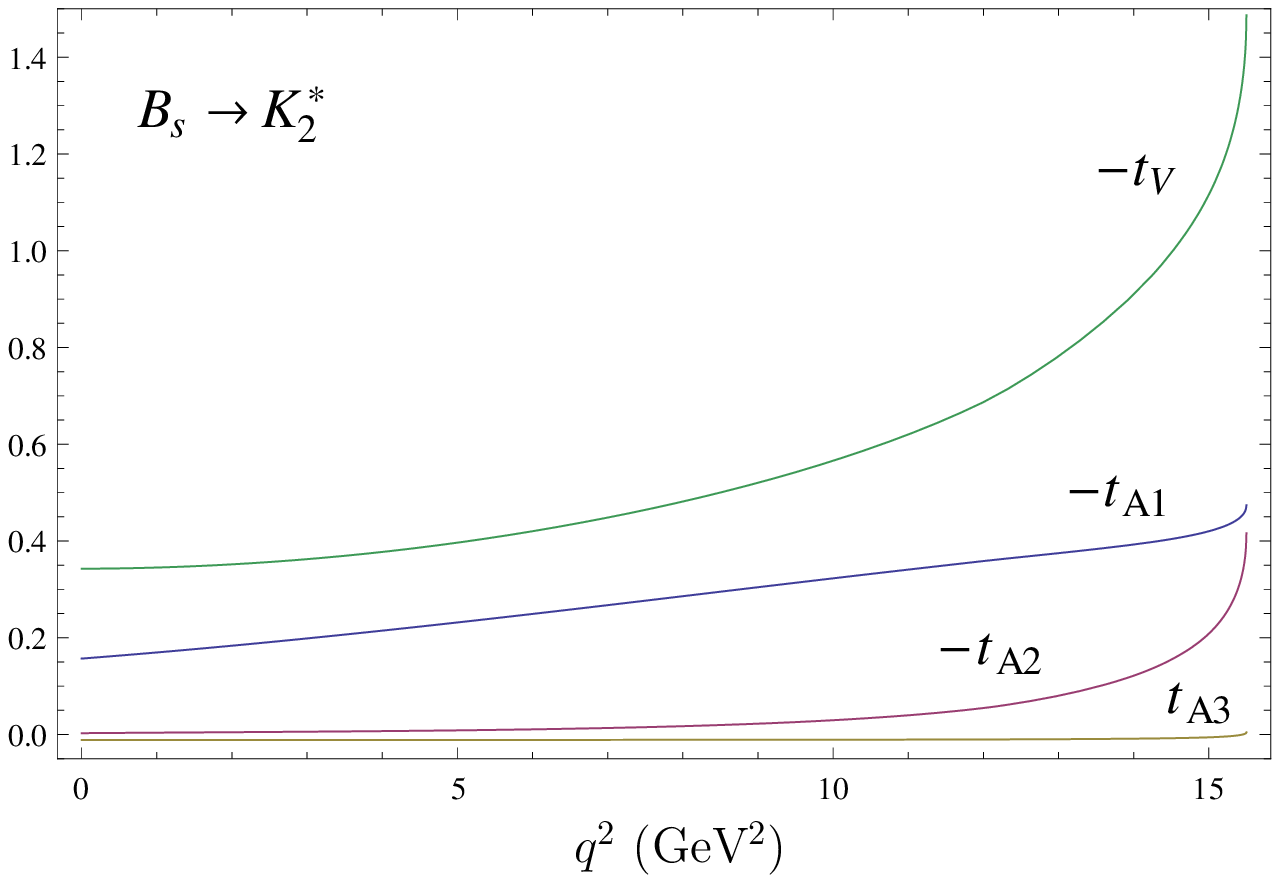}

  \caption{Form factors of the $B_s$ decays to the $P$--wave $K_{J}^{(*)}$ mesons.}
  \label{fig:ffbc}
\end{figure}

For the numerical
evaluations we again use the meson wave functions obtained in their mass spectra
calculations \cite{hlm,lm}. The calculated values of $B_s$ to
orbitally excited $K_{J}^{(*)}$ transition form factors at
maximum and zero recoil are give in Table~\ref{ffm}. The momentum
dependence of these form factors is shown in Fig.~\ref{fig:ffbc}. We
can estimate the errors in the calculated form factors to be
less than 10\%. They mainly originate from the evaluation of the
subleading contribution to the vertex function and uncertainties in
excited meson wave functions.

\begin{table}
\caption{Form factors of the weak $B_s$ decays to
  the $P$--wave $K_{J}^{(*)}$ mesons
  at $q^2=0$ and $q^2=q^2_{\rm max}\equiv(M_{B_s}-M_{K_{J}})^2$. }
\label{ffm}
\begin{ruledtabular}
\begin{tabular}{ccccccccccccccc}
&\multicolumn{2}{c}{$B_s\to K_{0}^*$}&\multicolumn{4}{c}{$B_s\to K_{1}(1400)$}&\multicolumn{4}{c}{$B_s\to K_{1}(1270)$}&\multicolumn{4}{c}{$B_s\to K_{2}^*$}\\
\cline{2-3} \cline{4-7} \cline{8-11} \cline{12-15}
$q^2$&$r_+$&$r_-$&$g_A$&$g_{V_1}$&$g_{V_2}$&$g_{V_3}$&$h_A$&$h_{V_1}$&$h_{V_2}$&$h_{V_3}$&$t_V$&$t_{A_1}$&$t_{A_2}$&$t_{A_3}$\\
\hline
0&0.27 &$-0.62$& $-0.33$&$-0.08$&$-0.16$&$-0.05$&$0.29$ &$0.08$ & $-0.14$&$0.42$
&$-0.34$&$-0.17$&$-0.01$&$-0.01$\\
$q^2_{\rm max}$&0.69 &$-1.59$& $-0.78$&0.25& $-0.75$&$0.02$&$0.98$ & $-0.16$ &$-0.03$ &$1.17$ 
&$-1.48$& $-0.48$&$-0.42$&$0.01$\\
\end{tabular}
\end{ruledtabular}
\end{table}

In Tables~\ref{compK0}-\ref{compK2} we compare our results for form
factors of the weak
$B_s$ decays to scalar $K_0^*$, axial vector $K_1$ and tensor $K_2^*$
mesons at  maximum recoil point $q^2=0$ with previous theoretical
calculations \cite{slh,ymz,wal,llww,hwfz,z,ykc,llw,w}.  Predictions are
mostly available for the weak $B_s$ decays to scalar $K_0^*$ mesons. Such
transitions were studied in the light-cone sum rules \cite{slh,wal},
QCD sum rules \cite{ymz,hwfz} and perturbative QCD approach
\cite{llww,z}. We find that our result for the form factor $r_+(0)$ is
consistent with Ref.~\cite{ymz}, but lower than in other calculations
\cite{slh,wal,llww,hwfz,z}. Contrary we predict larger
absolute value for the form factor $r_-(0)$ than in Refs.~\cite{slh,wal,hwfz}. 
Predictions for form factors of the weak $B_s$ decays to axial vector
$K_1$ mesons are compared in Table~\ref{compK1}. The light-cone sum
rule approach is used in Ref.~\cite{ykc}, while Ref.~\cite{llw}
employs perturbative QCD. From this table we find that our and these
theoretical approaches predict significantly different values of form
factors at maximum recoil. One of the origins can be the adopted values of
the mixing angle $\varphi$, defined in Eq.~(\ref{eq:mix}), of axial vector $K_1$
mesons. Note that in our model this angle $\varphi$ is explicitly calculated by
diagonalizing the mass matrix \cite{lm}, while in Refs.~\cite{ykc,llw}
different phenomenologically motivated values are used. The form
factors of the weak $B_s$ decays to tensor $K_2$ mesons were calculated in
Ref.~\cite{w} in the perturbative QCD approach. The obtained values of
the form factors $t_V(0)$ and $t_{A_1}(0)$ are slightly lower than in
our model. 

\begin{table}
\caption{Comparison of theoretical predictions for the form factors of 
the  weak $B_s\to  K^{*}_0$ transitions at maximum
  recoil point $q^2=0$.  }
\label{compK0}
\begin{ruledtabular}
\begin{tabular}{cccccccc}
$F(0)$& this
paper&\cite{slh}&\cite{ymz}&\cite{wal}&\cite{llww}&\cite{hwfz}&\cite{z}\\
\hline
$r_+(0)$&$0.27\pm0.03$&0.44 &$0.24\pm0.10$&$0.41^{+0.13}_{-0.07}$&
$0.56^{+0.16}_{-0.13}$& $0.39\pm0.04$&
$0.56^{+0.07}_{-0.10}$\\
$r_-(0)$&$-0.62\pm0.06$&$-0.44$ & &$-0.34^{+0.14}_{-0.09}$&
& $-0.25\pm0.05$
\end{tabular}
\end{ruledtabular}
\end{table}

\begin{table}
\caption{Same as in Table~\ref{compK0} but for the $B_s\to  K_1$ transitions.  }
\label{compK1}
\begin{ruledtabular}
\begin{tabular}{cccc}
$F(0)$& this
paper&\cite{ykc}&\cite{llw}\\
\hline
$g_A(0)$&$-0.33\pm0.03$&$-0.15^{+0.09}_{-0.07}$& $0.03\pm0.01$\\
$g_{V_1}(0)$&$-0.08\pm0.01$&$-0.11^{+0.07}_{-0.04}$& $0.11\pm0.07$\\
$h_A(0)$&$0.29\pm0.03$&$0.49\pm0.08$& $0.20\pm0.05$\\
$h_{V_1}(0)$&$0.08\pm0.01$&$0.38\pm0.06$& $0.87\pm0.25$
\end{tabular}
\end{ruledtabular}
\end{table}

\begin{table}
\caption{Same as in Table~\ref{compK0} but for the $B_s\to  K_2^*$ transition.  }
\label{compK2}
\begin{ruledtabular}
\begin{tabular}{ccc}
$F(0)$& this
paper&\cite{w}\\
\hline
$t_V(0)$&$-0.34\pm0.03$&$-0.18^{+0.05}_{-0.04}$\\
$t_{A_1}(0)$&$-0.17\pm0.02$&$-0.11^{+0.03}_{-0.02}$
\end{tabular}
\end{ruledtabular}
\end{table}

\section{Charmless semileptonic $B_s$ decays}
\label{sdbsk}
We can now apply the calculated form factors for the evaluation of the
semileptonic $B_s$ decays to ground state and orbitally excited $K$
mesons. The differential decay rate of the $B_s$ meson to a $K$
($K^{(*)}$ or $K^{(*)}_J$) meson can be expressed in the following form
\cite{iks}
\begin{equation}
  \label{eq:dgamma}
  \frac{d\Gamma(B_s\to Kl\bar\nu)}{dq^2}=\frac{G_F^2}{(2\pi)^3}
  |V_{ub}|^2
  \frac{\lambda^{1/2}(q^2-m_l^2)^2}{24M_{B_s}^3q^2} 
  \Biggl[(H_+H^{\dag}_++H_-H^{\dag}_-+H_0H^{\dag}_0)\left(1+\frac{m_l^2}{2q^2}\right)  +\frac{3m_l^2}{2q^2} H_tH^{\dag}_t\Biggr],
\end{equation}
where $G_F$ is the Fermi constant, $V_{ub}$ is the CKM matrix element, $\lambda\equiv
\lambda(M_{B}^2,M_F^2,q^2)=M_{B}^4+M_F^4+q^4-2(M_{B}^2M_F^2+M_F^2q^2+M_{B}^2q^2)$,
$m_l$ is the lepton mass. The transverse $H_\pm$, longitudinal
$H_0$ and time $H_t$  helicity
components of the hadronic tensor
 are defined through the
transition form factors calculated in the previous sections. The
corresponding relations for decays to ground and orbitally excited
mesons are given in Appendices C and E of Ref.~\cite{bsdecay}, respectively.

\subsection{Semileptonic $B_s$ decays to ground state $K^{(*)}$
  mesons}
\label{skgs}

\begin{figure}
  \centering
 \includegraphics[width=8cm]{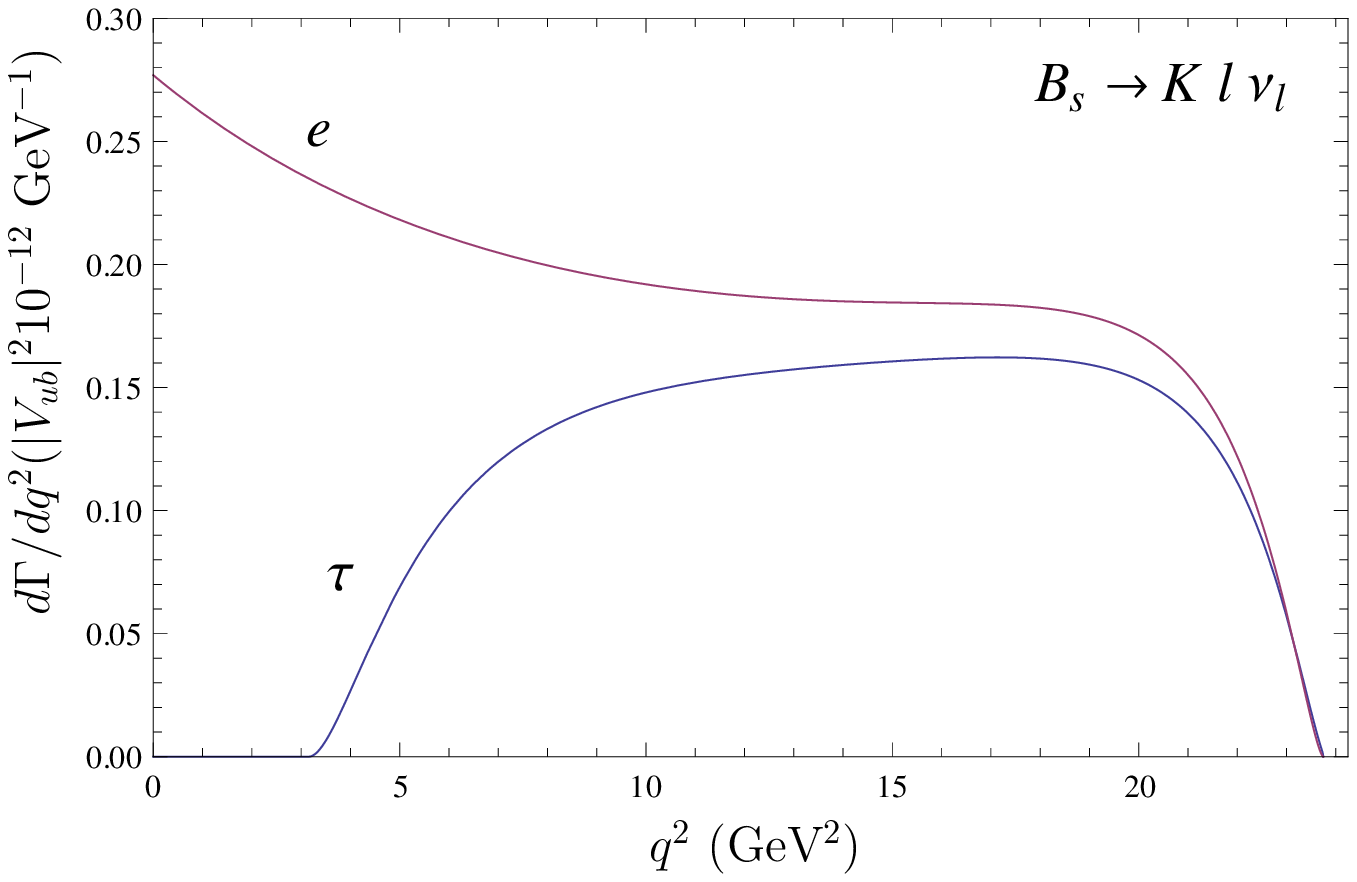}\ \
 \  \includegraphics[width=8cm]{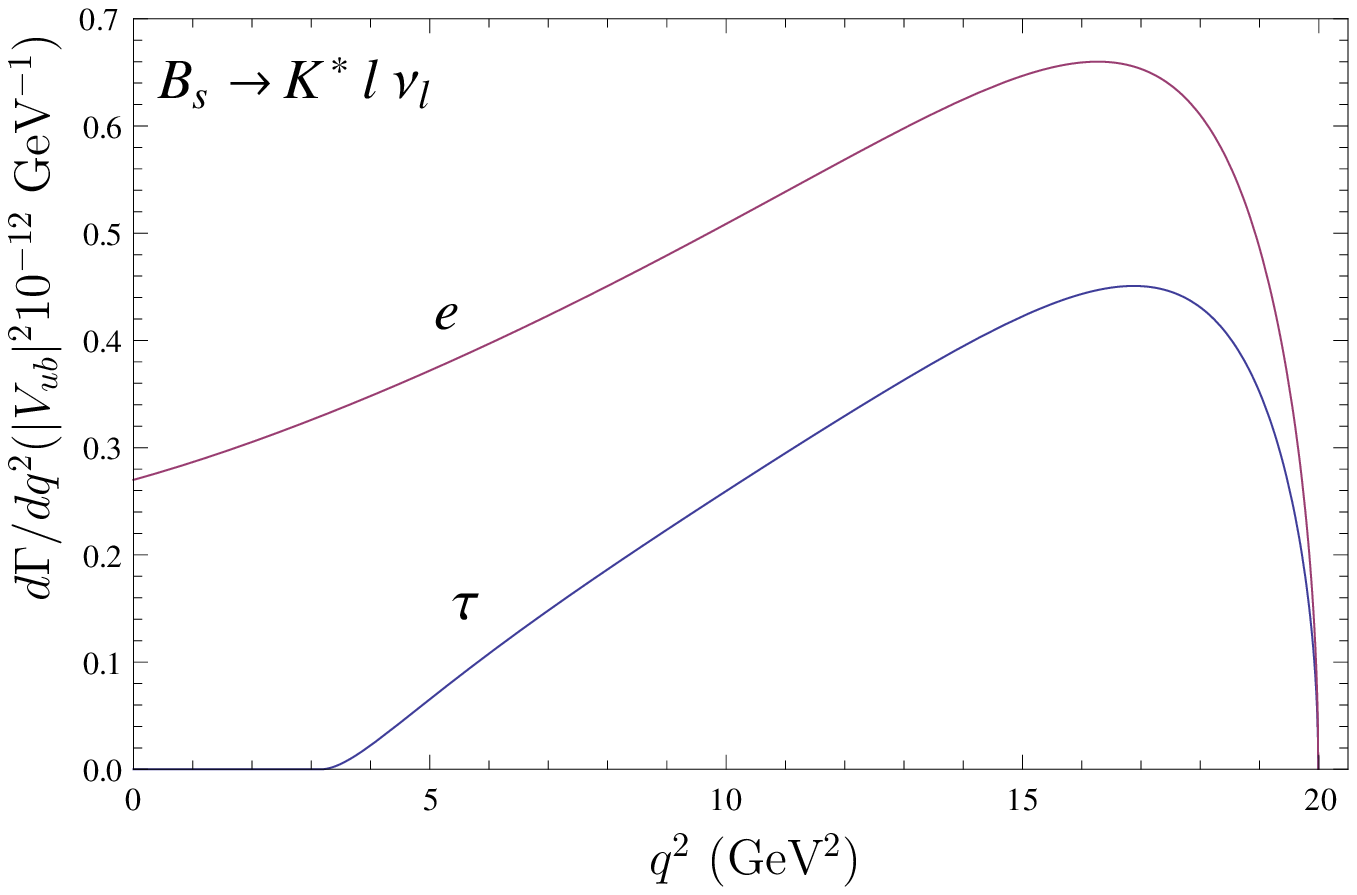}

  \caption{Predictions for the differential decay rates  of the  
    semileptonic $B_s\to K^{(*)}l\nu_l$ decays. }
  \label{fig:brbpr}
\end{figure}

First we calculate the rates of the semileptonic $B_s$ decays to ground state $K^{(*)}$
mesons. Substituting the form factors obtained in Sec.~\ref{sec:ffbsk}
in Eq.~(\ref{eq:dgamma}) we get the corresponding differential
decay rates. They are plotted in Fig.~\ref{fig:brbpr} both for decays
involving electron $e$ and $\tau$ lepton. Integrating these
differential decay rates over $q^2$ we find the total decay rates. In
calculations we use the value of the CKM matrix element
$|V_{ub}|=(4.05\pm 0.20)\times 10^{-3}$ found previously from the
comparison of the predictions of our model \cite{bdecays,btau} with
measured semileptonic $B\to\pi(\rho)l\nu_l$ decay rates. The
kinematical range accessible in the heavy-to-light $B_s\to K^{(*)}$ transitions is very
broad, that makes the knowledge of the $q^2$ dependence of the
form factors to be an important issue. Therefore, the explicit determination of
the momentum dependence of the weak decay form factors in the whole
$q^2$ range without any additional assumptions is an important
advantage of our model.  The 
calculated branching fractions of the semileptonic $B_s\to
K^{(*)}l\nu_l$ decays are presented in Table~\ref{compslbsk} in
comparison  with other theoretical predictions \cite{wx,wzz}. The
perturbative QCD factorization approach is used in Ref.~\cite{wx},
while in Ref.~\cite{wzz} light cone sum rules are employed. From the
comparison in Table~\ref{compslbsk} we see that all theoretical
predictions for the $B_s$ semileptonic branching fractions agree
within uncertainties. This is not surprising since these significantly
different approaches predict close values of the corresponding weak
form factors (see Table~\ref{compbpiff}).  

\begin{table}
\caption{Comparison of theoretical predictions for the branching
  fractions of semileptonic decays $B_s\to K^{(*)} l\nu_l$ (in $10^{-4}$).  }
\label{compslbsk}
\begin{ruledtabular}
\begin{tabular}{cccc}
Decay& this paper & \cite{wx}&\cite{wzz} \\
\hline
$B_s\to K e\nu_e$& $1.64\pm0.17$ & $1.27^{+0.49}_{-0.30}$  &$1.47\pm0.15$ \\ 
$B_s\to K\tau\nu_\tau$& $0.96\pm0.10$ &$0.778^{+0.268}_{-0.201}$   &$1.02\pm0.11$\\
$B_s\to K^*e\nu_e$& $3.47\pm0.35$ & & $2.91\pm0.26$ \\
$B_s\to K^*\tau\nu_\tau$& $1.67\pm0.17$ & &$1.58\pm0.13$\\
\end{tabular}
\end{ruledtabular}
\end{table}

We can use the calculated values of the semileptonic $B_s$ decay
branching fractions to obtain predictions for the ratios of such decay
involving $\tau$ lepton and electron or muon: $R(K)\equiv Br(B_s\to K
\tau\nu_\tau) /Br(B_s\to K e\nu_e)= 0.59\pm0.05$ and  $R(K^*)\equiv Br(B_s\to K^*
\tau\nu_\tau) /Br(B_s\to K^* e\nu_e)= 0.48\pm0.04$. The interest to
such ratios is stimulated by recently found deviations of experimental
data from theoretical predictions for the similar ratios for the
semileptonic $B$ decays to $D$ mesons $R(D^{(*)})=Br(B\to
D^{(*)}\tau\nu_\tau)/ Br(B\to D^{(*)}e\nu_e)$ (see, e.g., discussion in \cite{btau}
and references therein).
 
Summing up different contributions listed in Table~\ref{compslbsk} we
obtain the prediction for the total semileptonic $B_s$ decay branching
fraction to the ground state $K$ mesons to be $Br(B_s\to K^{(*)}
e\nu_e)=(5.11\pm0.51)\times 10^{-4}$ and $Br(B_s\to K^{(*)}
\tau\nu_\tau)=(2.63\pm0.26)\times 10^{-4}$.

\subsection{Semileptonic $B_s$ decays to orbitally excited $K_{J}^{(*)}$
  mesons}
\label{skoe}

\begin{figure}
  \centering
 \includegraphics[width=8cm]{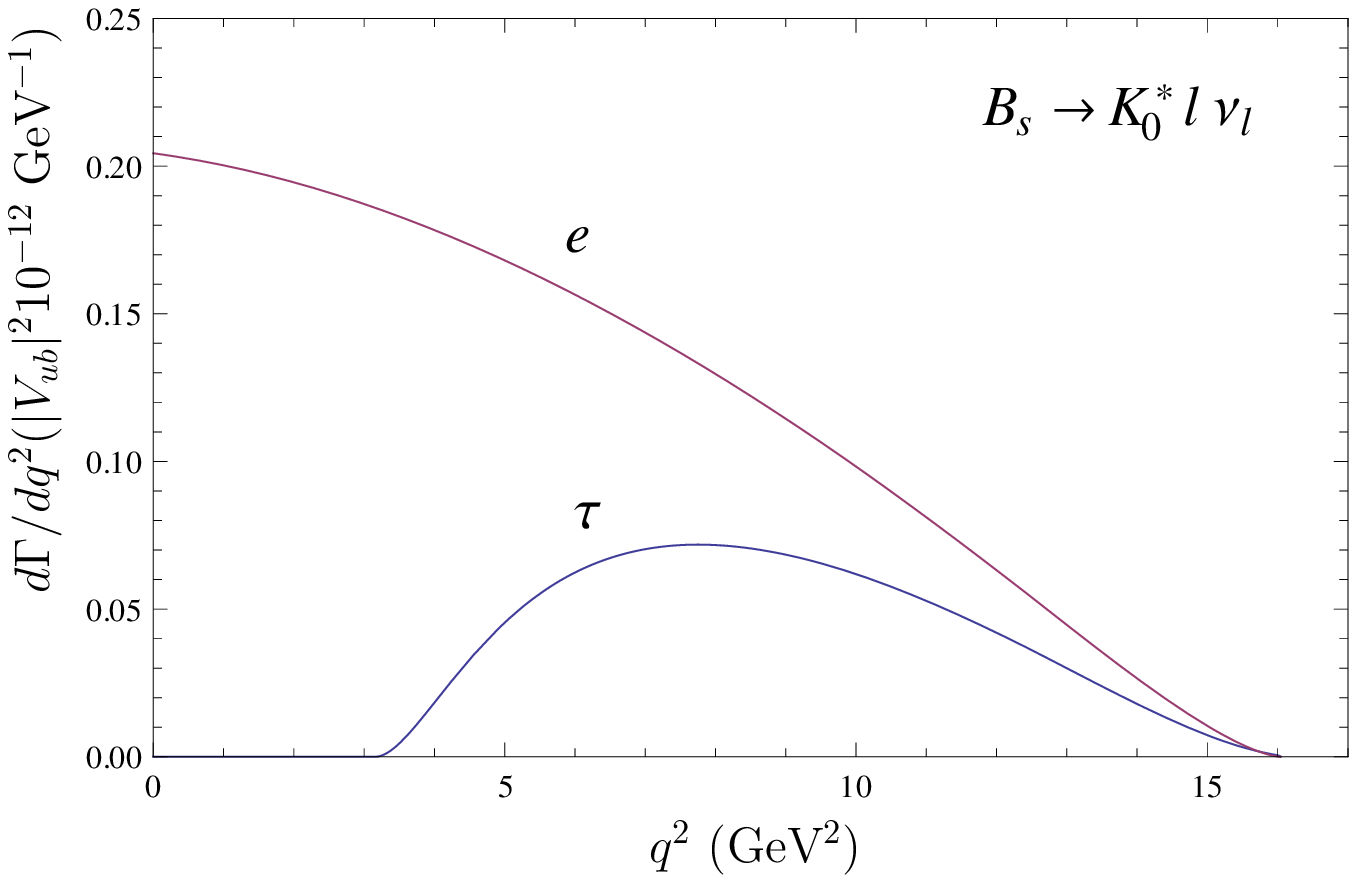}\ \
 \  \includegraphics[width=8cm]{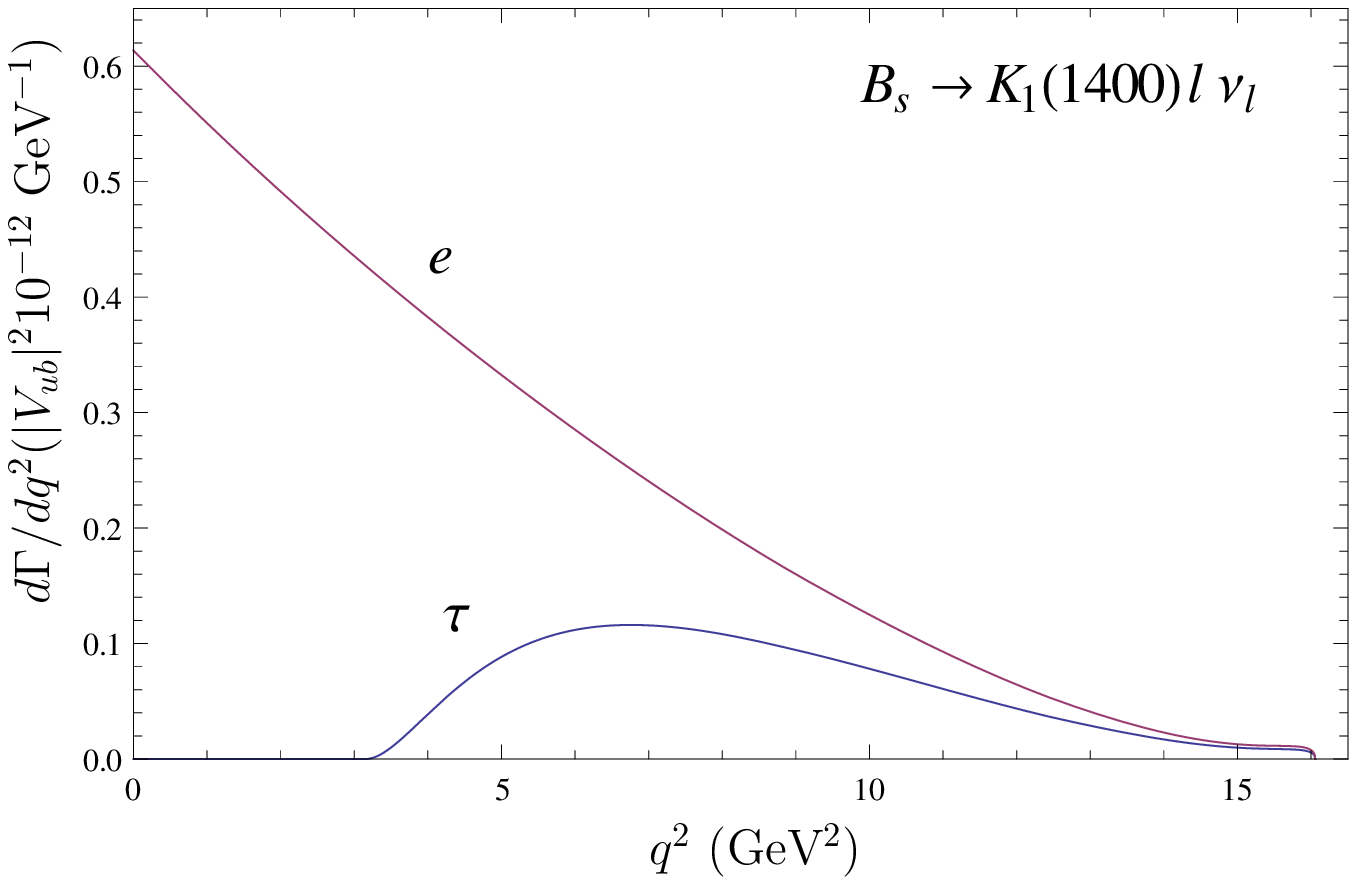}\\
\includegraphics[width=8cm]{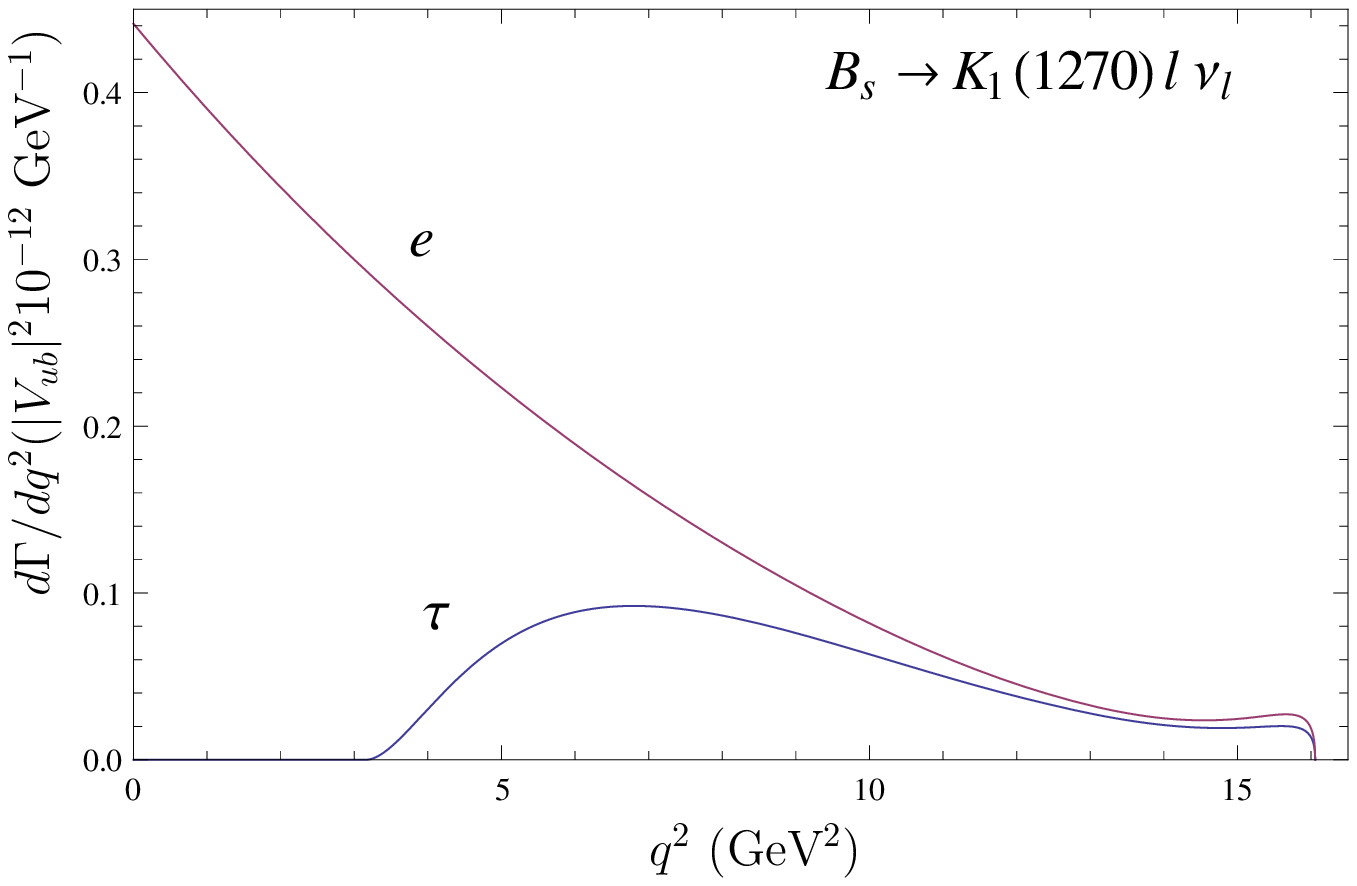}\ \
 \  \includegraphics[width=8cm]{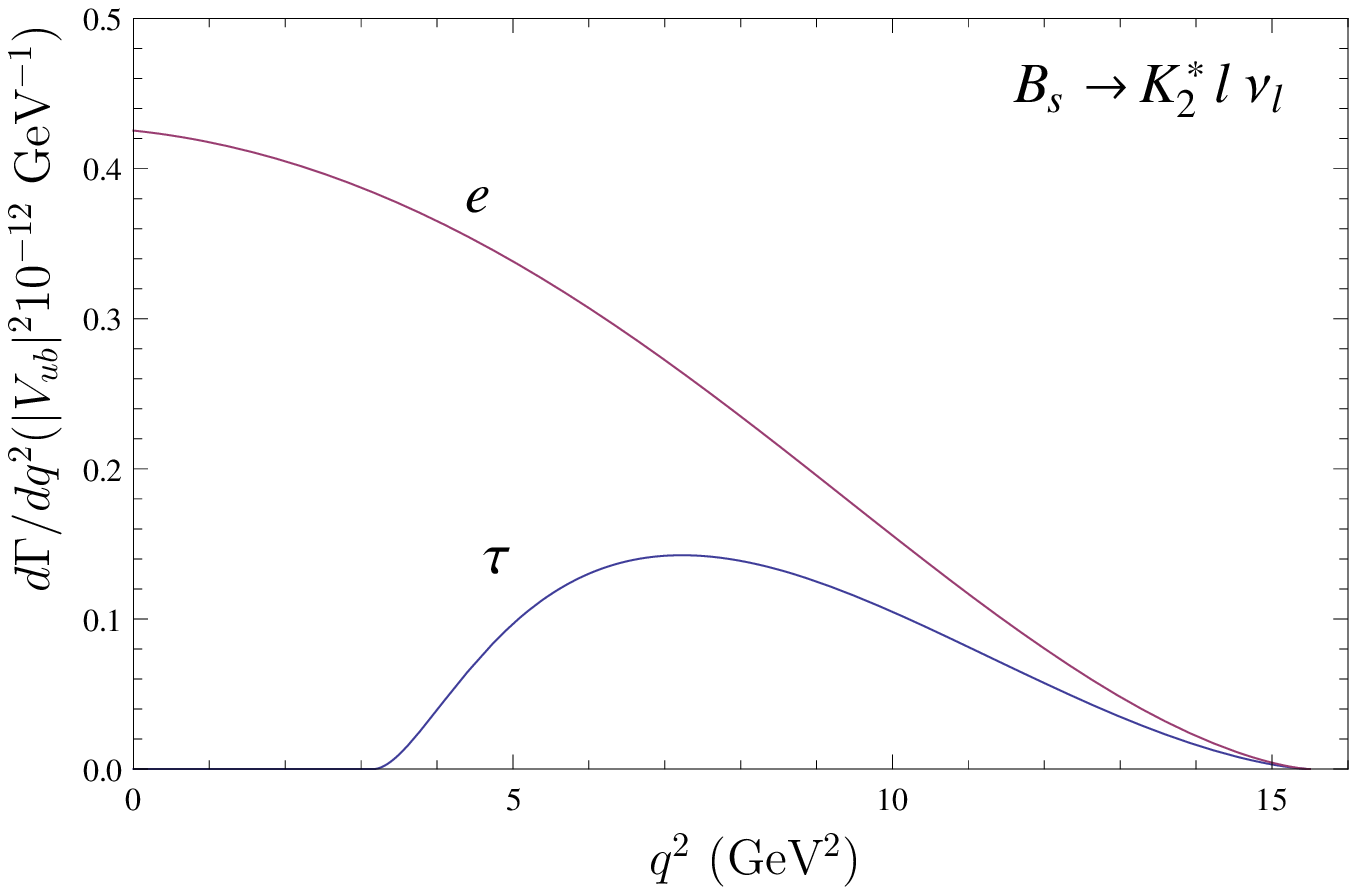}
\caption{Predictions for the differential decay rates  of the
  semileptonic $B\to K_{J}^{(*)}l\nu_l$ 
     decays. }
  \label{fig:brboe}
\end{figure}

Now we calculate the branching fractions of the semileptonic $B_s$
decays to orbitally excited $K_{J}^{(*)}$ mesons. We substitute the
form factors obtained in Sec.~\ref{sec:fforbexc} in the expression for
the differential decay rate (\ref{eq:dgamma}). The resulting decay
rates are plotted in Fig.~\ref{fig:brboe}. Integration over $q^2$
gives the total semileptonic  $B_s\to K_J^{(*)} l\nu_l$ branching
fractions which are given in Table~\ref{comphlff}. We see that our
model predicts close values (about $1\times10^{-4}$) for all
semileptonic $B_s$ branching fractions to the first orbitally excited $K_{J}^{(*)}$
mesons. Indeed, the difference between branching fractions is less than
a factor of 2. This result is in contradiction to the dominance of specific
modes (by more than a factor of 4) in the heavy-to-heavy semileptonic $B\to D_J^{(*)}
l\nu_l$ and  $B_s\to D_{sJ}^{(*)} l\nu_l$ decays
\cite{bexdecay,bsdecay}, but it is consistent with predictions for
the corresponding
heavy-to-light semileptonic $B$ decays to orbitally excited light
mesons \cite{borbldecay}. The above mentioned suppression of some
heavy-to-heavy decay channels to orbitally excited heavy mesons was
mostly pronounced in the heavy quark limit and then slightly reduced by the
heavy quark mass corrections which are found to be large
\cite{bexdecay}. Thus our result once again indicates 
that the $s$ quark cannot be treated as a heavy one and should be
considered to be light instead, as we always did in our calculations.  

\begin{table}
\caption{Comparison of theoretical predictions for the branching
  fractions of semileptonic decays  $B_s\to K_J^{(*)} l\nu_l$ (in $10^{-4}$).  }
\label{comphlff}
\begin{ruledtabular}
\begin{tabular}{cccccccc}
Decay& this paper &\cite{ymz}& \cite{wal}&\cite{llww}  &\cite{ykc} &\cite{llw}&\cite{w}
 \\
\hline
$B_s\to K^*_0 e\nu_e$& $0.71\pm0.14$&  $0.36^{+0.38}_{-0.24}$& $1.3^{+1.3}_{-0.4}$& $2.45^{+1.77}_{-1.05}$ &  
& &  \\
$B_s\to K^*_0 \tau\nu_\tau$& $0.21\pm0.04$&& $0.52^{+0.57}_{-0.18}$& $1.09^{+0.82}_{-0.47}$   &  
& &  \\
$B_s\to K_1(1270) e\nu_e$& $1.41\pm0.28$  & & & &  $4.53^{+1.67}_{-2.05}$&
$5.75^{+3.49}_{-2.89}$\\
$B_s\to K_1(1270) \tau\nu_\tau$& $0.30\pm0.06$ & &&& & $2.62^{+1.58}_{-1.31}$
\\
$B_s\to K_1(1400) e\nu_e$& $0.97\pm0.20$ & & &  &$3.86^{+1.43}_{-1.75}$&
$0.03^{+0.05}_{-0.02}$ \\
$B_s\to K_1(1400) \tau\nu_\tau$& $0.25\pm0.05$ & & & && $0.01^{+0.02}_{-0.01}$
\\
$B_s\to K^*_2 e\nu_e$& $1.33\pm0.27$ & & & & &&$0.73^{+0.48}_{-0.33}$  \\
$B_s\to K^*_2 \tau\nu_\tau$& $0.36\pm0.07$ & & & &&& $0.25^{+0.17}_{-0.12}$ \\
 \end{tabular}
\end{ruledtabular}
\end{table}

In Table~\ref{comphlff} we compare our predictions for the
semileptonic $B_s$ branching fractions to orbitally excited $
K_J^{(*)}$ mesons with previous calculations
\cite{ymz,wal,llww,ykc,llw,w}. The consideration in
Ref.~\cite{ymz} is based on QCD sum rules. The light cone sum rules are
used in Refs.~\cite{wal,ykc}, while Refs.~\cite{llww,llw,w} employ the
perturbative QCD approach. Reasonable agreement between our results
and other predictions \cite{ymz,wal,w} is
observed for the semileptonic $B_s$ decays to the scalar and tensor $K$
mesons. The values  of Ref.~\cite{llww} are almost a factor 3
higher. For the semileptonic $B_s$ decays to axial vector $K$
mesons predictions are significantly different even within rather
large errors. Therefore experimental measurement of these decay
branching fractions can help to discriminate between theoretical
approaches.

For the total semileptonic $B_s$ decays to first orbital excitations
of $K$ mesons we get $Br(B_s\to K_J^{(*)}
e\nu_e)=(4.4\pm0.9)\times 10^{-4}$ and $Br(B_s\to K_J^{(*)}
\tau\nu_\tau)=(1.1\pm0.2)\times 10^{-4}$. These values are close to
the ones found for the semileptonic $B_s$ decays to ground state $K$
mesons in Sec.~\ref{skgs}. The similar pattern of the branching
fraction dependence on the excitation of a final meson was previously found for
the heavy-to-light semileptonic $B$ decays \cite{borbldecay}. On the
other hand,
for the heavy-to-heavy semileptonic $B\to D$ and $B_s\to D_s$ decays
the pronounced hierarchy, where the decay branching fractions rapidly
decrease with the growing excitation of the final heavy ($D$ or $D_s$) meson,
is observed \cite{bexdecay,bsdecay}.

\section{Charmless nonleptonic decays of $B_s$ mesons}\label{nl}

We further apply calculated $B_s$ decays form factors to the
evaluation of the charmless nonleptonic decays of the $B_s$ meson. In the
discussion we
closely follow our previous consideration of nonleptonic decays of $B$
mesons to ground and orbitally excited states of light mesons in the
factorization approximation \cite{nonl,borbldecay}. To
simplify the problem we limit our analysis to the calculation of the
decay processes  dominated by the tree 
diagrams. For such decays the matrix elements of the effective weak
Hamiltonian $H_{\rm eff}$, governing nonleptonic decays $B_s\to K^{-,0}M^{+,0}$,
where $M$ is a light ($\pi$ or $\rho$) meson, can be approximated by the product of
one-particle transition amplitudes
 \begin{eqnarray}
  \label{eq:fact}
  \langle K^-M^+|H_{\rm eff}|B_s\rangle &\approx& \frac{G_F}{\sqrt{2}}
V_{ub}^*V_{ud}
a_1^{\rm eff}\langle K^-|(\bar b u)_{V-A}|B_s\rangle \langle M^+|(\bar 
d u)_{V-A}|0\rangle\cr 
\langle K^0M^0|H_{\rm eff}|B_s\rangle &\approx& \frac{G_F}{\sqrt{2}}
V_{ub}^*V_{ud} a_2^{\rm eff}\langle K^0|(\bar b d)_{V-A}|B_s\rangle \langle M^0|(\bar u
u)_{V-A}|0\rangle,
\end{eqnarray}
with
$$a_1^{\rm
  eff}=a_1-\frac{V_{tb}^*V_{td}}{V_{ub}^*V_{ud}}[a_4+a_{10}+r_q(a_6+a_8)],$$ 
$$a_2^{\rm eff}=a_2-\frac{V_{tb}^*V_{td}}{V_{ub}^*V_{ud}}\left[-a_4\mp
\frac32 a_{7}+\frac32 a_{9}+\frac12 a_{10}-r_q(2a_6-a_8)\right],$$ 
where terms in square brackets result from the penguin
contributions, which are small numerically; minus and plus correspond
to $M^0=\pi^0$ and $M^0=\rho^0$,
respectively. The quantities $a_{2n-1}=c_{2n-1}+c_{2n}/N_c$ and
$a_{2n}=c_{2n}+c_{2n-1}/N_c$ ($n=1,2\dots$ and $N_c$ is the number of colors) are combinations of
the Wilson coefficients $c_i$, which we take from Ref.~\cite{wc}, and $r_q$  can be found, e.g., in Ref.~\cite{rss}. 

The matrix element of the weak current
$J^W_\mu$ between vacuum and a final 
pseudoscalar ($P$) or vector ($V$) meson can be parametrized by the decay
constants $f_{P,V}$
\begin{equation}
\langle P|\bar q_1 \gamma^\mu\gamma_5 q_2|0\rangle=if_Pp^\mu_P, \qquad
\langle V|\bar q_1\gamma_\mu q_2|0\rangle=\epsilon_\mu M_Vf_V.
\end{equation}
As a result the corresponding nonleptonic matrix element factorizes in
the product of the weak $B_s\to K$ decay form factors, which were
calculated in the  previous sections, and decay constants. The
pseudoscalar $f_P$ and vector $f_V$ decay constants of light and heavy
mesons were calculated within our model in Ref.~\cite{fpconst}. Their
values are in agreement with the available experimental data \cite{pdg}. For the
calculations  we use the following values of the decay constants:
$f_\pi=0.131$~GeV, $f_\rho=0.208$~GeV. The relevant CKM
matrix elements are $|V_{ud}|=0.975$,  $|V_{td}|=0.0087$,
$|V_{tb}|=0.999$ \cite{pdg}.

In Table~\ref{compnl} we present the predictions for the branching
fractions of the charmless nonleptonic $B_s$  decays to ground state
$K$ mesons obtained in the factorization approximation with our model
form factors. There we also give results of other theoretical approaches
\cite{akllsww,bn,wz,cc,swyz} and available experimental data
\cite{pdg}. Calculations in Ref.~\cite{akllsww} are done in
perturbative QCD approach. QCD factorization is employed in
Refs.~\cite{bn,cc}. The authors of Ref.~\cite{wz} use soft-collinear
effective theory, while considerations in Ref.~\cite{swyz} are based
on an approximate six-quark  effective Hamiltonian. Experimentally
only the $B_s\to K^-\pi^+$   branching fraction was measured \cite{pdg}
and for the $B_s\to K^{*0}\rho^0$ decay upper limit is available. All
theoretical predictions agree with data.  In fact all theoretical
results for decays into charged $K^{(*)-}$ and $\pi^+(\rho^+)$ mesons
are consistent within rather large error bars, while for decays
involving neutral mesons deviations are larger.

\begin{table}
\caption{Comparison of various predictions for the branching fractions
  of the charmless nonleptonic
  $B_s$  decays to ground state $K$ mesons with
  experiment (in $10^{-6}$).  }
\label{compnl}
\begin{ruledtabular}
\begin{tabular}{cccccccc}
Decay&this paper&\cite{akllsww} &\cite{bn}&\cite{wz}&\cite{cc} & \cite{swyz}&Experiment \cite{pdg} \\
\hline
$B_s\to K^-\pi^+$& $8.7\pm2.7$ & $7.6^{+3.3}_{-2.5}$&$10.2^{+6.0}_{-5.2}$&$4.9\pm1.8$ &$5.3^{+0.5}_{-0.9}$  & $7.1^{+3.3}_{-1.8}$&
$5.3\pm1.0$  \\ 
$B_s\to K^-\rho^+$& $24.0\pm7.2$&  $17.8^{+7.9}_{-5.9}$&$24.5^{+15.2}_{-12.9}$&$10.2\pm1.0$  & $14.7^{+1.7}_{-2.3}$&
$17.6^{+8.2}_{-4.6}$  \\
$B_s\to K^{*-}\pi^+$& $8.6\pm2.6$&  $7.6^{+3.0}_{-2.3}$& $8.7^{+5.9}_{-4.9}$ &$6.6\pm0.7$ & $7.8^{+0.6}_{-0.9}$&
 $7.2^{+5.6}_{-2.3}$  \\
$B_s\to K^{*-}\rho^+$& $25.4\pm7.6$&
$20.9^{+8.4}_{-6.5}$&$25.2^{+4.9}_{-3.5}$&   &
$21.6^{+1.6}_{-3.2}$&$21.0^{+13.5}_{-6.5}$  \\
$B_s\to K^0\pi^0$& $0.25\pm0.08$ & $0.16^{+0.11}_{-0.06}$&$0.49^{+0.63}_{-0.35}$&$0.76\pm0.41$ &$1.7^{+2.7}_{-0.9}$  & $1.1^{+0.7}_{-0.3}$& \\ 
$B_s\to K^0\rho^0$& $0.67\pm0.20$&  $0.08^{+0.07}_{-0.04}$&$0.61^{+1.26}_{-0.60}$&$0.81\pm0.09$  & $1.9^{+3.2}_{-1.1}$&
$0.6^{+0.3}_{-0.2}$  \\
$B_s\to K^{*0}\pi^0$& $0.24\pm0.07$&  $0.07^{+0.05}_{-0.03}$& $0.25^{+0.46}_{-0.22}$ &$1.07\pm0.19$ & $0.89^{+1.16}_{-0.49}$&
 $0.3^{+0.2}_{-0.2}$  \\
$B_s\to K^{*0}\rho^0$& $0.71\pm0.21$&
$0.33^{+0.17}_{-0.11}$&$1.5^{+3.3}_{-1.5}$&   &
$1.3^{+2.6}_{-0.7}$&$1.0^{+0.4}_{-0.3}$ &$<767$ \\
\end{tabular}
\end{ruledtabular}
\end{table}

We apply the same factorization approach for the calculation of the
nonleptonic decays to orbitally excited $K^{(*)}_J$ mesons. Using form
factors obtained in Sec.~\ref{sec:fforbexc} we get predictions for the
nonleptonic branching fractions and present them in
Table~\ref{compnlexc} in comparison with previous estimates
\cite{z,sdv}. A few predictions are available for selected modes. In
Ref.~\cite{z} decay  $B_s\to K_{0}^{*-}\rho^+$ was 
considered within the perturbative QCD factorization approach. The
obtained central value of this decay branching fraction is almost a
factor of 4 larger than our result. This is the consequence of a 2 times
larger form factor $r_+(0)$ in Ref.~\cite{z} than in our
model (see Table~\ref{compK0}). Decays involving the tensor $K_{2}^{*-}$
meson were considered in Ref.~\cite{sdv} within the
Isgur-Scora-Grinstein-Wise II model. The predictions are
approximately a factor of 2 lower than our central values of
branching fractions.  

\begin{table}
\caption{Branching fractions of the nonleptonic
  $B_s$  decays to orbitally excited  $K^{(*)}_J$ mesons (in $10^{-6}$).  }
\label{compnlexc}
\begin{ruledtabular}
\begin{tabular}{cccc}
Decay& this paper&\cite{z}&\cite{sdv}\\
\hline
$B_s\to K_{0}^{*-}\pi^+$& $9.6\pm3.8$&   \\ 
$B_s\to K_{0}^{*-}\rho^+$& $27\pm10$& $108^{+34}_{-31}$\\
$B_s\to K_{1}(1270)\pi^+$&$29\pm12$&  \\
$B_s\to K_{1}(1270)\rho^+$& $76\pm30$&  \\
$B_s\to K_{1}^{-}(1400)\pi^+$& $21\pm8$&  \\
$B_s\to K_{1}^{-}(1400)\rho^+$& $54\pm21$&  \\
$B_s\to K_{2}^{*-}\pi^+$& $17\pm6$&  &7.8 \\ 
$B_s\to K_{2}^{*-}\rho^+$& $47\pm18$&& 23\\
\end{tabular}
\end{ruledtabular}
\end{table} 

\section{Conclusions}
\label{sec:concl}

The form factors of the $B_s$ weak decays to strange mesons were
calculated in the framework of the relativistic quark model based on
the quasipotential approach. Decays both to the ground $K^{(*)}$ and
to orbitally
excited  $K^{(*)}_J$ mesons were considered. The form factors were
determined as the overlap integrals of the related meson wave functions, found
in the previous meson mass spectrum calculations \cite{hlm,lm}, in the
whole broad kinematical range without additional assumptions about
their $q^2$ dependence. The important relativistic effects, such as
transformations of the meson wave functions from the rest to a
moving reference frame and contributions of the intermediate
negative-energy states, were consistently taken into account. 

We used these form factors for the evaluation of the charmless
semileptonic $B_s$ decay rates to ground state and orbitally excited
$K$ mesons. It was found that total branching fractions of semileptonic
$B_s$ decays to ground and first orbitally excited $K$ mesons have
close values about $5\times 10^{-4}$. Summing up these contributions
we get $(9.5\pm1.0)\times 10^{-4}$. This value is almost 2 orders of magnitude
lower than our prediction for the corresponding sum of branching
fractions of the semileptonic $B_s$ to $D_s$
mesons \cite{bsdecay} as it was expected from the ratio of CKM matrix
elements $|V_{ub}|$ and $|V_{cb}|$. Therefore the total semileptonic $B_s$ decays
branching fraction is dominated by the decays to $D_s$ mesons and in our model is equal 
to $(10.3\pm1.0)\%$  in good agreement with the experimental value
$Br(B_s\to X e\nu_e)_{\rm Exp.}=(9.5\pm2.7)\%$ \cite{pdg}. 
The obtained predictions for charmless semileptonic $B_s$ decays were
compared with previous calculations. It was found that
different theoretical approaches yield close values of branching fractions for decays to
ground state $K^{(*)}$ mesons agreeing within uncertainties, while the ones
for the decays to orbitally excited $K^{(*)}_J$ mesons differ
significantly from each other. The latter observation can
help to discriminate between theoretical models.  

Charmless two-body nonleptonic $B_s$ decays dominated by tree diagrams
were considered. The factorization approximation was used which
allowed to express the decay matrix elements as the product of weak
transition matrix elements and decay constants. The branching fractions of the
nonleptonic decays to ground state $K^{(*)}$ or orbitally excited $K^{(*)}_J$ meson and
pion or $\rho$ meson, both charged and 
neutral, were calculated. The obtained results were
confronted with previous theoretical predictions and experimental
data, which are available for only few of the considered decays. Good
agreement with data and other evaluations is found for the nonleptonic
decays to the ground state $K^{(*)}$ meson and a light meson, while again for decays involving
the orbitally excited  $K^{(*)}_J$ mesons significant disagreement
between predictions of different approaches is observed.

\acknowledgments
The authors are grateful to D. Ebert, M.~A.~Ivanov, V. A. Matveev,
M. M\"uller-Preussker and V. I. Savrin  
for  useful discussions.
This work was supported in part by the {\it Russian
Foundation for Basic Research} under Grant No.12-02-00053-a.

\end{document}